\documentclass[%
 reprint,
nofootinbib,
 amsmath,amssymb,
 aps,
]{revtex4-2}

\usepackage{aas_macros}
\usepackage{graphicx}
\usepackage{dcolumn}
\usepackage{bm}

\usepackage{chemformula}

\usepackage[dvipsnames]{xcolor}

\definecolor{BerlinU1}{HTML}{62AD2D}
\definecolor{BerlinU2}{HTML}{E94D10}
\definecolor{BerlinU3}{HTML}{00A192}
\definecolor{BerlinU9}{HTML}{F18800}

\usepackage[colorlinks=true]{hyperref}
\hypersetup{
    colorlinks = true,
    linkcolor = BerlinU9,
    urlcolor = BerlinU9,
    linkbordercolor = {white},
    citecolor = BerlinU9,
    }

\newcommand{\ord}[1]{\mathcal{O} \left( #1 \right)}
\newcommand{\D}{\mathcal{D}}
\allowdisplaybreaks[4]

\usepackage{orcidlink}
\begin{document}

\title{Numerical simulations of low-mass black hole--neutron star mergers \\ with moderate eccentricities}

\author{Ivan \surname{Markin}\,\orcidlink{0000-0001-5731-1633}$^{1}$}
\email{ivan.markin@uni-potsdam.de}
\author{Gonzalo \surname{Morrás}\, \orcidlink{0000-0002-9977-8546}$^{2}$ }
\author{Tim \surname{Dietrich}\,\orcidlink{0000-0003-2374-307X}$^{1,2}$}

\affiliation{${}^1$ Institut f\"ur Physik und Astronomie, Universit\"at Potsdam, Haus 28, Karl-Liebknecht-Str. 24/25, 14476, Potsdam, Germany}
\affiliation{${}^2$ Max Planck Institute for Gravitational Physics (Albert Einstein Institute), Am M\"uhlenberg 1, Potsdam 14476, Germany}

\date{\today}

\begin{abstract}
A confident detection of orbital eccentricity in gravitational-wave signals from black hole-neutron star systems would be a strong indicator of their dynamical formation origin. 
While eccentric waveform models have recently been developed, their validity for analyzing black hole–neutron star signals has yet to be tested against numerical-relativity simulations. Building upon our previous study of quasicircular black hole-neutron star mergers, we perform simulations of a subset of these systems with orbital eccentricities around five times larger than those inferred for GW200105\_162426. Additionally, we perform simulations with magnetized matter to assess whether trapped $g$-mode oscillations persist in strongly magnetized disks.
Regarding the emitted gravitational-wave signal, we find that the effective one-body waveform model SEOBNRv5EHM shows reasonable agreement unless tidal contributions become important, and the inspiral-only post-Newtonian models pyEFPE/pyEFPEHM show larger disagreement in the late inspiral. For the equal-mass system, which has the largest effective tidal deformability, eccentric tidal corrections have a small but noticeable impact on the phase, $\sim0.1$~rad.
We find an almost negligible impact of eccentricity on the properties of the remnant black hole, and differences of up to a factor of two in the remnant disk mass. Trapped $g$ modes in the accretion disks have the same frequency regardless of the eccentricity, confirming that they depend mainly on the properties of the remnant black hole. We find that when accretion disks are strongly magnetized, they become less dense and have less pronounced structure, and that the trapped $g$ modes previously identified in the unmagnetized cases are efficiently damped.
\end{abstract}

\maketitle

\section{Introduction}
Multiple gravitational-wave (GW) signals of black hole-neutron star (BHNS) mergers have been detected to date, most likely, GW200105\_162426 (hereafter GW200105)~\cite{LIGOScientific:2021qlt}, GW200115\_042309~\cite{LIGOScientific:2021qlt}, and GW230529\_181500~\cite{LIGOScientific:2024elc}. Typically, BHNS systems can form primarily via evolution of isolated stellar binaries with a common envelope phase~\cite{Paczynski1976,Ostriker1976,Bethe:1998bn,Belczynski:2001uc} or through dynamical formation in dense stellar environments, e.g., globular and nuclear clusters~\cite{Samsing:2017xmd,Rodriguez:2017pec,Fragione:2018yrb,Zevin:2018kzq}, and young stellar clusters~\cite{Banerjee:2016ths,Banerjee:2017mgr,DiCarlo:2019pmf}.
While in most formation channels high natal eccentricities can be achieved ($e \gtrsim 0.1$)~\cite{Dhurkunde:2023qoe}, these eccentricities will be efficiently radiated away during inspiral~\cite{Peters:1964zz,Hinder:2007qu} to the values of $e \ll 0.1$ in case of isolated binary formation or formation in globular clusters. 

On the other hand, binaries originating through dynamical formation pathways, such as in hierarchical triples or nuclear clusters, may retain high eccentricities ($e \gtrsim 0.1$) when they enter the detector
band at $20~\text{Hz}$. Therefore, a detection of orbital eccentricity of a compact binary merger by ground-based detectors would suggest its dynamical formation origin. Also, the eccentricity can be introduced through the presence of a tertiary object~\cite{Stegmann:2025clo}.

Until recently, most GW analyses employed only quasicircular (QC) waveform models, i.e., models that neglect orbital eccentricity~\cite{LIGOScientific:2018mvr,LIGOScientific:2020ibl,LIGOScientific:2021usb,KAGRA:2021vkt,LIGOScientific:2025hdt,LIGOScientific:2025zdk,LIGOScientific:2026sit,LIGOScientific:2026ifv}.
However, employing QC waveform models for the analysis of eccentric signals is known to cause biases in the parameter estimation, cf. Ref.~\cite{Huez:2025npe} for low-mass systems and Ref.~\cite{RoyChowdhury:2026xgb} for stellar-mass binary black hole (BBH) systems. 
Also, employing QC waveform models can lead to misinterpreting the orbital eccentricity as very high individual spins~\cite{Chandra:2026voe}.

Recently, following developments in eccentric waveform models, multiple reanalyses of GW events using eccentric waveform models were performed, e.g.,~Refs.~\cite{Morras:2025xfu,Planas:2025plq,Jan:2025fps,Kacanja:2025kpr}. Those studies provide evidence for a nonzero eccentricity of $e_\text{20Hz}\sim 0.15$ for GW200105, suggesting the dynamical origin of the source. Moreover, these studies found support for zero component spins, indicating that eccentric BHNS systems might have no detectable spins and spin precession. Given its large mass ratio and negligible BH spin, the neutron star (NS) is expected to directly plunge into the black hole (BH), not producing bright electromagnetic transients. Indeed, no electromagnetic counterpart was identified~\cite{LIGOScientific:2021qlt,Anand:2020eyg}, leaving no further clues regarding its origin. In addition, the inferred eccentricity appears to be strongly prior-driven~\cite{Clarke:2026cuw}, casting doubt on the strength of the nonzero eccentricity claims. 

\begin{table*}[!ht]
    \centering
    \setlength{\tabcolsep}{6pt}
    \begin{tabular}{l|ccccc|cccc|ccc}
        Name & $m_\mathrm{BH}$ & $m_\mathrm{NS}$ & $Q$ & $M$ & $\mathcal{M}_c$ & EOS &  $\mathcal{C}_\mathrm{NS}$ & $\Lambda_\mathrm{NS}$ & $\tilde{\Lambda}$  & $e_\text{ID}$ & $e_\mathrm{GW}$ & $e_\mathrm{GW}^{20\mathrm{Hz}}$  \\
        \hline

        \verb|Q1_SLy_e0.1| &  1.4  & 1.396 & $1$ & 2.796 & 1.2170 &  SLy & 0.1799  & 313 & 156 & $0.1$  & 0.12 & $0.81$ \\

        \verb|Q2_SLy_e0.05| &  2.8  & 1.397 & $2$ & 4.197 & 1.7015 &  SLy & 0.1800 & 311 & 39 & $0.05$ & 0.06 & $0.64$ \\
        \verb|Q2_SLy_e0.1| &  2.8  &  1.397 & $2$ & 4.197 & 1.7015 &  SLy &  0.1800 & 311 & 39 & $0.1$  & 0.14 &  $0.77$\\
        
        \verb|Q3_SLy_e0.1| &  4.2  & 1.397 & $3$ & 5.597 & 2.0487 &  SLy & 0.1800 & 311 & 14  & $0.1$  & 0.15 & $0.73$ \\

        \hline
        
        \verb|Q2_DD2_e0.05| &  2.8  & 1.402 & $2$ & 4.202 & 1.7047 &  DD2 & 0.1568 & 678 & 86  & $0.05$  & 0.06 & $0.64$ \\
        \verb|Q2_DD2_e0.1| &  2.8  & 1.402 & $2$ & 4.202 & 1.7047 &  DD2 & 0.1568 & 679 & 86  & $0.1$  & 0.14 & $0.77$ \\
    \end{tabular}
    \caption{Binary parameters for the simulated configurations.
    \textit{First column:} name of the configuration. \textit{First group of columns:} $m_\mathrm{BH}$ is the Christodoulou mass of the BH, $m_\mathrm{NS}$ the ADM mass of the NS in isolation, $Q=m_\mathrm{BH}/m_\mathrm{NS} \geq 1$ is the mass ratio, $M$ is the total mass of the system defined as the sum of the individual gravitational masses, $\mathcal{M}_c$ is the chirp mass. \textit{Second group of columns:} The employed EOS, the NS compactness $\mathcal{C}_{\mathrm{NS}}$, the quadrupolar tidal deformability of the NS $\Lambda_\mathrm{NS}$, and the effective tidal deformability of the binary $\tilde{\Lambda}$. \textit{Third group of columns:} $e_\mathrm{ID}$ is the fiducial orbital eccentricity set in the initial data, $e_\mathrm{GW}$ is the eccentricity calculated from the waveform at the reference dimensionless frequency $f_\mathrm{ref} M = 0.0087$, $e_\mathrm{GW}^{20\mathrm{Hz}}$ is $e_\mathrm{GW}$ rewinded (evolved backwards in time) to 20~Hz using Peters formula~\cite{Peters:1964zz}.}
    \label{table:configurations}
\end{table*}

While most of the BHNS modelling in numerical relativity (NR) focused on QC binaries, a handful of simulations focused on eccentric BHNS systems~\cite{Stephens:2011as,East:2011xa,East:2015yea,Jan:2025fps}. To our knowledge, no study has systematically simulated longer BHNS inspirals with moderate eccentricities. With an increasing number of GW models, including eccentricity~\cite{Ramos-Buades:2021adz,Gamboa:2024hli,Gonzalez:2025xba,Gamboa:2026jht,Morras:2025nlp,Morras:2026fho}, some of them also including tidal effects~\cite{Gonzalez:2025xba,Morras:2026fho}, it is instrumental to populate the parameter space with long NR waveforms of eccentric BHNS systems. With such numerical waveforms available, it would be possible to assess the current standing of the eccentric BHNS waveform models and potentially calibrate them.

To fill this gap, we perform numerical-relativity simulations of low-mass BHNS mergers with eccentricities $e^\mathrm{GW}_{20\mathrm{Hz}}\in [0.64, 0.81]$ and nonspinning components. Building upon our previous study~\cite{Markin:2026eyc}, we select the configurations with mass ratios $Q = M_\mathrm{BH} / M_\mathrm{NS}  \in \{1, 2, 3\}$ at a fixed NS mass, and vary their eccentricity. Additionally, we perform simulations with magnetized NSs to study the effects of strongly magnetized disks on the presence of trapped $g$-mode oscillations in the accretion disk.

The paper is structured as follows.
In Sec.~\ref{section:setup}, we introduce the simulated configurations, describe the construction of eccentric initial data, and the numerical setup for the evolution.
In Sec.~\ref{section:results}, we describe the results of these simulations, such as numerical waveforms, remnant properties, and accretion disks.
In Sec.~\ref{sec:magnetic_fields}, we describe the simulations with magnetic fields, and discuss the effects of strong magnetic fields on the dynamics of the accretion disks and trapped $g$-mode oscillations, and comment on the magnetization of the disks and formation of a polar funnel.
We conclude in Sec.~\ref{section:conclusion}.

Throughout the paper, we use geometrized units, i.e., where $G=c=\mathrm{M_\odot}=1$, unless stated otherwise. For the magnetic field quantities, we use Heaviside–Lorentz units, i.e., 
magnetic energy is defined as $E_\mathrm{mag}=b_a b^a/2$, with $b^a$ being the four-vector of magnetic field strength in the fluid frame.

\section{Numerical setup}
\label{section:setup}

We select two representative equations of state (EOSs), one soft and one stiff, SLy~\cite{Douchin:2001sv} and DD2~\cite{Typel:2009sy}, respectively. For both EOSs, we use their piecewise-polytropic representation: the parameters from Ref.~\cite{Read:2008iy} for SLy, and a piecewise-polytropic fit to the EOS table from CompOSE~\cite{Typel:2013rza,Oertel:2016bki,CompOSECoreTeam:2022ddl} for DD2. As in Ref.~\cite{Markin:2026eyc}, the DD2 initial data has a mismatch between the StellarCollapse DD2~\cite{OConnor:2009iuz} used for the initial data construction and the CompOSE DD2~\cite{CompOSECoreTeam:2022ddl} used for the evolution. Therefore, the gravitational masses and tidal deformabilities in Tab.~\ref{table:configurations} are recalculated using the baryonic masses on the numerical grid at the beginning of the evolution.

The configurations and their parameters are listed in Tab.~\ref{table:configurations}. 

\subsection{Initial data}
We construct initial data for the configurations using \textsc{FUKA}~\cite{Papenfort:2021hod}, which is based on the \textsc{Kadath} spectral library~\cite{Grandclement:2009ju}. \textsc{FUKA} solves the Einstein constraint equations using the extended conformal thin sandwich (XCTS) formulation~\cite{York:1998hy,Pfeiffer:2002iy}.

Aiming to make the eccentric waveforms of comparable length to their QC counterparts studied in our previous work, Ref.~\cite{Markin:2026eyc}, we equate the mean motions of the binaries to those of the corresponding QC cases.

\textsc{FUKA} is designed to produce QC binaries, and, therefore, allows adjustments of the initial radial infall and the angular velocities to perform eccentricity reduction. For that reason, we leverage this ability to adjust these parameters to introduce eccentricity instead of reducing it. For simplicity of the initial data construction, we consider the binaries initially at apoapsis. We start by finding the initial angular velocity at apoapsis $\Omega_{\text{apo},0}$ for the desired eccentricity from the initial angular velocity $\Omega_{\text{QC},0}$ of the QC configurations of Ref.~\cite{Markin:2026eyc} keeping the mean motion fixed,
\begin{equation}
    \Omega_{\text{apo},0} = \frac{\sqrt{1-e_\mathrm{ID}^2}}{(1+e_\mathrm{ID})^2} \Omega_{\text{QC},0}.
\end{equation}

\noindent Then, we rescale the initial separation of the QC cases, $d_\text{QC}$, accordingly to obtain the initial separation for the eccentric case
\begin{equation}
    d_\text{apo} = (1+e_\mathrm{ID}) d_\text{QC}.
\end{equation}
\noindent Finally, we determine the radial velocity using the 3.5~post-Newtonian (PN) expression for generalized Kepler's third law, Eqs.~(372a) and (372b) of Ref.~\cite{Blanchet:2013haa}, to complete the procedure.

In general relativity, eccentricity does not have a unique definition. To bring both NR simulations and waveform models onto the same footing, a definition of eccentricity measured from the gravitational waveform was introduced in Refs.~\cite{Shaikh:2023ypz,Shaikh:2025tae}.
For our simulations, we calculate the eccentricity, $e_\mathrm{GW}$, measured shortly after the beginning of the simulations, at the reference dimensionless frequency $f_\mathrm{ref} M = 0.0087$. To allow the eccentricities to be more easily compared with the literature, we introduce $e_\mathrm{GW}^\mathrm{20Hz}$, which is $e_\mathrm{GW}$ rewinded (evolved backwards in time) to $20$~Hz using Peters formula~\cite{Peters:1964zz}. We list all these eccentricities in Tab.~\ref{table:configurations}.

Because all configurations here are set to start at apoapsis, the initial true anomaly $\nu$ is close to $\pi$. As shown in Ref.~\cite{Nee:2025zdy}, changes in the true anomaly in the initial data primarily shift merger time within a $10M$-wide range, where $M$ is the total mass of the binary. For the heaviest binary considered here with $Q=3$ (prefixed with \verb|SLy_Q3|), that would produce maximum changes on the order of $\sim 0.28$~ms in case of $\nu=0$. In our case, we assume that the initial true anomaly, while not being exactly $\pi$, remains close to $\pi$ across all configurations. Therefore, such deviation in the true anomaly has an otherwise negligible impact on the inspiral duration. However, as we discuss below, the true anomaly at the time of the tidal disruption may determine its outcome.

\subsection{Evolution}
We evolve the systems with the general-relativistic (magneto)hydrodynamics code \textsc{BAM}~\cite{Bruegmann:2006ulg,Thierfelder:2011yi,Dietrich:2015iva,Bernuzzi:2016pie,Dietrich:2018phi}. There, we employ the conformal Z4 reformulation~\cite{Hilditch:2012fp} (Z4c) of Einstein field equations for the evolution of the spacetime, coupled with the moving-puncture gauge~\cite{Campanelli:2005dd,Baker:2005vv}. The (magneto)hydrodynamics is solved for in the finite-volume Valencia formulation~\cite{Marti:1991wi,Banyuls:1997zz,Anton:2005gi,Font:2008fka} with high-resolution shock-capturing scheme consisting of the local Lax-Friedrichs (LLF) Riemann solver and the \mbox{WENO-Z} reconstruction~\cite{2008JCoPh.227.3191B} for primitive variables. 

Within \textsc{BAM}, the Cartesian grid consists of nested refinement levels, with each finer level having twice the resolution of its parent. The three outermost levels are fixed, and the other levels are allowed to move. Each moving level may contain subgrids that fully cover the compact objects at all times. In cases where these subgrids overlap, they are merged into one box without changing the spacing. 
We denote the resolutions by the number of grid points $n$ along spatial dimensions at the finest level, and refer to them as $\mathrm{R}n$.
We perform all unmagnetized simulations at three distinct resolutions: R96, R144, and R192. We refer to the results of the highest-resolution (R192) runs as the main results throughout the paper and use the other resolutions for the convergence studies in Appendix~\ref{section:convergence}.
We perform time integration using a fourth-order Runge-Kutta scheme with Courant-Friedrichs-Lewy factor set to $0.25$, and employ the Berger-Colella scheme~\cite{Berger:1989a} for local time stepping, and maintain the conservation of the baryonic mass, energy, and momentum across the refinement levels~\cite{Berger:1989a,Dietrich:2015iva}. For the evolution of the magnetic fields, \textsc{BAM} employs a hyperbolic divergence cleaning scheme~\cite{DEDNER2002645,Liebling:2010bn,Penner:2010px,Mosta:2013gwu} with direct evolution of magnetic field components. The details of the implementation are described in Ref.~\cite{Neuweiler:2024jae}. For the conservative-to-primitive recovery, we employ the \textsc{RePrimAnd} algorithm~\cite{Kastaun:2020uxr}. The magnetized cases are evolved only at the highest resolution, R192.

\section{Results}
\label{section:results}

\subsection{Numerical waveforms}

\begin{figure*}[htp]
    \centering
    \includegraphics[width=1\linewidth]{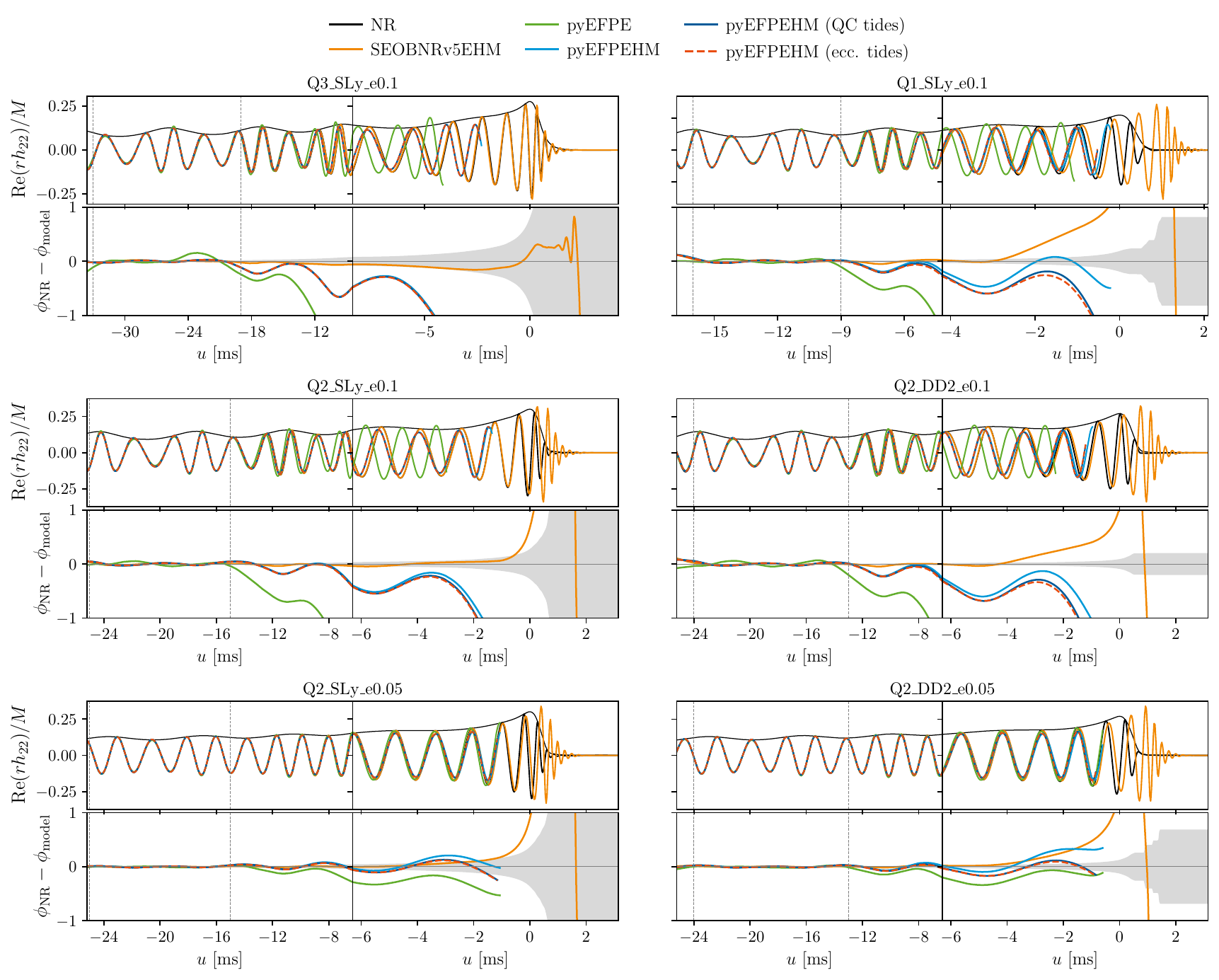}
    \caption{Comparison of the NR waveforms with eccentric waveform models for each configuration. \textit{Top panels}: the real part of the total-mass- and distance-rescaled strain $\mathrm{Re}(rh_\mathrm{22})/M$. The vertical gray dashed lines delineate the alignment window. \textit{Bottom panels}: phase difference between the numerical waveform and waveform models. The gray band marks the NR phase error calculated as a monotonically increasing phase difference between the highest (R192) and the medium (R144) resolutions.}
    \label{fig:waveform_models}
\end{figure*}

We employ two eccentric waveform families: the effective-one-body model SEOBNRv5EHM~\cite{Gamboa:2024hli,Mihaylov:2023bkc}, and the inspiral-only PN waveform models pyEFPE~\cite{Morras:2025nlp} and pyEFPEHM~\cite{Morras:2026fho} with two additional variants employing adiabatic tides with either QC or eccentricity-dependent coefficients. For the implementation details of eccentric tidal contributions~\cite{Henry:2025uta,Henry:2026bqh} in pyEFPEHM, cf.~Appendix~\ref{sec:appendix:ecc_tides}.

For pyEFPEHM, higher-order PN corrections are included in both the phase and the amplitude compared to pyEFPE. However, adding higher PN orders does not necessarily guarantee an increase in the waveform model accuracy, and it depends on at which order the PN series is truncated~\cite{Sun:2026lfd}.

We align each waveform model to the NR data by minimizing the time-domain mismatch of the $(2,2)$ mode over a fixed window in retarded time $u \in [u_0, u_1]$. The windows are chosen to include around $1.5$ radial periods at the beginning of the signal.

In order to accelerate the minimization procedure and to reduce the dimensionality of the search space, we perform the following procedure. Writing the real part of the $(2,2)$ mode NR waveform as $s(u)=\mathrm{Re}\!\left[h_{22}^\mathrm{NR}(u)\right]$ and the model polarizations as $h_+$ and $h_\times$, the template is
\begin{equation}
    h(u) = \mathrm{Re}\!\left[\kappa\,(h_+ + \mathrm{i} h_\times)(u-\delta t)\right]=\kappa_+ \,h_+ + \kappa_\times\,h_\times ,
    \label{eq:htemplate}
\end{equation}
with $\kappa=\kappa_+ - \mathrm{i} \kappa_\times$, so that a single complex coefficient $\kappa$ captures the overall amplitude, a constant phase shift, and the polarization angle. Using the unweighted time-domain inner product over the window
\begin{equation}
    \langle x,y\rangle=\int_{u_0}^{u_1}x(u)\,y(u)\,\mathrm{d}u,
    \label{eq:innerprod}
\end{equation}
the time-domain mismatch is then given by
\begin{align}
    \mathcal{MM}_\mathrm{TD} &= 1 - \max_{\substack{e_0,\,\ell_0,\,\delta t,\\ \kappa }} \frac{\langle h, s \rangle}{\sqrt{\langle h, h \rangle \langle s, s \rangle}} \, .
    \label{eq:mismatch_def} 
\end{align}
The minimization over polarization can then be analytically performed. To this end, we introduce
\begin{equation}
    v_i = \langle s, h_i \rangle, \quad G_{i j} = \langle h_i, h_j \rangle, \quad i,j \in \{+, \times\} \, ,
    \label{eq:viGij}
\end{equation}
where $\bm{v}$ is a two-component vector, and $\bm{\hat{G}}$ is a $2 \times 2$ symmetric and positive-definite matrix, and therefore can be used as a metric to define the scalar product
\begin{equation}
    (\bm{a}, \bm{b}) = \bm{a}^T \bm{\hat{G}} \bm{b} \, .
    \label{eq:scalarprod}
\end{equation}
Writing $\bm{\kappa} \equiv (\kappa_+, \kappa_\times)^T$, the overlap can then be rewritten as
\begin{align}
    \frac{\langle h, s \rangle}{\sqrt{\langle h, h \rangle \langle s, s \rangle}} = \frac{1}{\sqrt{\langle s, s \rangle}}\frac{(\bm{\kappa},\bm{\hat{G}}^{-1} \bm{v})}{\sqrt{(\bm{\kappa},\bm{\kappa})}} \, .
    \label{eq:mismatch_scalarprod} 
\end{align}
From Eq.~\eqref{eq:mismatch_scalarprod} and the Cauchy--Schwarz inequality for $(\cdot,\cdot)$, it follows that the maximum overlap will be reached when $\bm{\kappa}$ is parallel to $\bm{\hat{G}}^{-1}\bm{v}$ (i.e., $\bm{\kappa}_\mathrm{opt} = \alpha \, \bm{\hat{G}}^{-1}\bm{v}$ with $\alpha>0$). In such a case, the mismatch of Eq.~\eqref{eq:mismatch_def} takes the following value
\begin{align}
    \mathcal{MM}_\mathrm{TD} &= 1 - \max_{e_0,\,\ell_0,\,\delta t} \frac{\sqrt{\bm{v}^T \bm{\hat{G}}^{-1} \bm{v}}}{\sqrt{\langle s, s \rangle}} \, .
    \label{eq:mismatch_opt} 
\end{align}
An analogous maximization is derived in the frequency domain in Ref.~\cite{Harry:2016ijz} using a complex inner product and maximizing over a constant phase shift independently of the polarization angle. The remaining maximizations are numerical. The time shift $\delta t$ is scanned in steps of the sampling interval, by shifting the waveform without recomputing it, and then refining it below the sampling interval. The maximization over the reference eccentricity $e_0$ and mean anomaly $\ell_0$ is performed using the differential evolution algorithm~\cite{Storn:1997uea,Scipy:2020}.

Many eccentric waveform models have their own internally defined eccentricity parameter, which is known to differ from the one measured from the waveform, cf.~Sec.~3.3 of Ref.~\cite{Vijaykumar:2024piy}. Therefore, we treat the reference eccentricity $e_0$ and mean anomaly $\ell_0$ of the waveform model as free parameters over which we minimize the mismatch. We list the optimal eccentricity values from these alignments in Appendix~\ref{sec:alignment_values}.

In Fig.~\ref{fig:waveform_models}, we show the resulting numerical waveforms and the corresponding waveform model signals aligned to them.

First, we consider \verb|Q3_SLy_e0.1|, whose waveform is expected to closely resemble a BBH waveform because the NS plunges directly into the BH. This is reflected in the good agreement with the NR-calibrated BBH model SEOBNRv5EHM. Throughout the duration of the waveform, the dephasing largely remains within the error band of the NR data.
On the other hand, the inspiral-only pyEFPE/pyEFPEHM models show stronger disagreements both within and after the alignment window. The worst agreement among these models is shown by pyEFPE. This is expected, as pyEFPE is less physically complete, including lower PN orders for the orbital phasing.
Conversely, pyEFPEHM, both with and without tidal contributions, substantially improves the agreement, although the dephasing still oscillates outside the NR error band. All of the pyEFPEHM variants, i.e., without and with tidal contributions, essentially overlap, confirming that tidal contributions are small for this mass ratio.

In contrast, for the equal-mass setup \verb|Q1_SLy_e0.1|, tidal effects are noticeably larger. Consequently, most models show a larger disagreement with the NR waveform in the late inspiral. For SEOBNRv5EHM, the dephasing moves out of the NR error band at about six GW cycles before merger. As the tidal corrections are not included in the model, the NR waveform has an earlier merger time than the BBH model SEOBNRv5EHM and shows a characteristic shutoff instead of a ringdown.
The pyEFPEHM model shows oscillating dephasing from the NR data described above. Inclusion of tidal contributions to pyEFPEHM introduces a noticeable phase difference that continuously grows from $\sim 5$~ms before merger, and reaches $\sim 0.5$~rad in the late inspiral.
Of special interest is the inclusion of contributions for eccentric tides--they introduce additional phase contribution on the order of $0.1$~rad in the late inspiral. This highlights their growing importance in more eccentric systems with equal mass ratios and, by extension, larger effective tidal deformability $\tilde{\Lambda}$.

For the $Q=2$ cases with higher eccentricities, \verb|Q2_SLy_e0.1| and \verb|Q2_DD2_e0.1|, the waveform models show similar dephasing trends. The dephasing of SEOBNRv5EHM leaves the NR error band about six GW cycles before merger for \verb|Q2_DD2_e0.1|, whereas for \verb|Q2_SLy_e0.1| it remains within the band almost until merger, consistent with the smaller $\tilde{\Lambda}$. As before, pyEFPE has the worst agreement and starts dephasing right after the alignment window. pyEFPEHM agrees with the NR waveform better, but shows strongly oscillating dephasing with a larger amplitude than the NR error band. The oscillations of the dephasing correlate with the radial period, suggesting that some eccentric contributions remain unaccounted for in pyEFPEHM. For the pyEFPEHM model with inclusion of tidal contributions, the phase evolves consistently faster than for the BBH case. This is consistent with the expectation that tidal effects accelerate the coalescence. For the stiffer DD2 EOS, the phase deviations are larger than for SLy, which is expected as the NS in the DD2 case has larger tidal deformability.

For the $Q=2$ cases with lower eccentricities, \verb|Q2_SLy_e0.05| and \verb|Q2_DD2_e0.05|, the agreement improves, with the largest improvements visible for pyEFPE/pyEFPEHM models. Notably, the pyEFPE waveforms become substantially closer to the pyEFPEHM ones. Yet, the oscillating dephasings of pyEFPE/pyEFPEHM still correlate with the radial period, suggesting unaccounted eccentric effects as in the $Q=2$ cases with higher eccentricity.

\subsection{Remnant properties}
We list remnant properties in Tab.~\ref{table:remnant_properties}, and discuss them separately below. The baryonic mass of the remnant and the properties of the remnant BH are evaluated at $10$~ms after merger.
We augment the table with magnetized cases, and discuss them separately in Sec.~\ref{sec:magnetic_fields}. 

\begin{table}[!tp]
    \centering
    \begin{tabular}{l|r|r|cc}
        Name & $M_{\mathrm{ej}}^{\mathrm{dyn}}$ & $M_{\mathrm{disk}}$  & $M_{\mathrm{BH}}^{\mathrm{rem}}$  &  $\chi_{\mathrm{BH}}^{\mathrm{rem}}$ \\
        \hline

        \verb|Q2_DD2_e0| &  $3.01 \times 10^{-3}$    & $8.21 \times 10^{-2}$ & $4.034$ & $0.67$ \\
        \verb|Q2_DD2_e0.05| &  $2.72 \times 10^{-3}$ &  $7.98 \times 10^{-2}$ & $4.037$ & $0.67$ \\
        \verb|Q2_DD2_e0.1| &  $3.25 \times 10^{-3}$  &  $8.50 \times 10^{-2}$  & $4.033$ & $0.67$ \\
        
        \hline

        \verb|Q2_SLy_e0|    & $< 1 \times 10^{-4}$ &  $1.30 \times 10^{-2}$ & $4.077$ & $0.67$ \\
        \verb|Q2_SLy_e0.05| & $< 1 \times 10^{-4}$ & $1.71 \times 10^{-2}$ & $4.075$ & $0.67$ \\
        \verb|Q2_SLy_e0.1|  & $< 1 \times 10^{-4}$ &  $2.51 \times 10^{-2}$ & $4.069$ & $0.67$ \\
        \verb|Q2_SLy_e0.1_poloidal|  & $7.73 \times 10^{-4}$ &  $3.08 \times 10^{-2}$ & $4.065$ & $0.67$ \\
        \verb|Q2_SLy_e0.1_toroidal|  & $1.75 \times 10^{-3}$ &  $2.71 \times 10^{-2}$  & $4.071$ & $0.67$ \\
        \hline

        \verb|Q1_SLy_e0|   &  $< 1 \times 10^{-4}$ & $2.89 \times 10^{-3}$ & $2.735$ & $0.81$ \\
        \verb|Q1_SLy_e0.1| &  $< 1 \times 10^{-4}$ & $1.86 \times 10^{-3}$  & $2.736$ & $0.81$ \\

        \hline

        \verb|Q3_SLy_e0|   & $< 1 \times 10^{-4}$ &  $6.17 \times 10^{-4}$ & $5.453$ &  $0.55$ \\
        \verb|Q3_SLy_e0.1| & $< 1 \times 10^{-4}$ &  $3.71 \times 10^{-4}$ & $5.452$ & $0.55$ \\
    \end{tabular}
    \caption{Dynamical ejecta mass  $M_{\mathrm{ej}}^{\mathrm{dyn}}$, baryonic mass of the disk $M_{\mathrm{disk}}$, remnant BH mass $M_{\mathrm{BH}}^{\mathrm{rem}}$ and spin $\chi_{\mathrm{BH}}^{\mathrm{rem}}$ for the binary configurations. The table is augmented with the QC simulations of Ref.~\cite{Markin:2026eyc}, suffixed with "\_e0", and magnetized simulations suffixed with the name of the magnetic field topology employed.}
    \label{table:remnant_properties}
\end{table}

Out of all unmangetized cases, only the \verb|Q2_DD2| cases have a non-negligible mass of dynamical ejecta. There is no clear dependency of the dynamical ejecta mass and disk mass on the eccentricity. For the \verb|Q2_SLy| cases, the disk mass increases with eccentricity, and decreases with eccentricity for the \verb|Q1_SLy| and \verb|Q3_SLy| cases. We suggest that the mass outcomes primarily depend on the phase of the radial motion (true anomaly) at the time of the tidal disruption: the NS could be disrupted earlier than it would have been in the QC cases if the disruption radius is reached by the NS due to a closer passage (periapsis). Consequently, this will adjust the impact parameter at the merger, which is likely the quantity responsible for regulating the merger outcomes in eccentric neutron star binaries~\cite{Neuweiler:2025lte}.

The mass and the spin of the remnant BH are unaffected by the initial eccentricity considered here. These results are consistent with the conclusions of Ref.~\cite{Hinder:2007qu} for BBH mergers, where both the mass and spin of the remnant BH remain unchanged as long as the eccentricities in the late inspiral are not significantly large, $e \lesssim 0.5$. The vanishing imprint of high eccentricities on the remnant BH properties was also found in the simulations of BHNS mergers in Ref.~\cite{Jan:2025fps}.

\subsection{Accretion and trapped $g$ modes}
\label{section:accretion}
\begin{figure}[htp]
    \centering
    \includegraphics[width=1\linewidth]{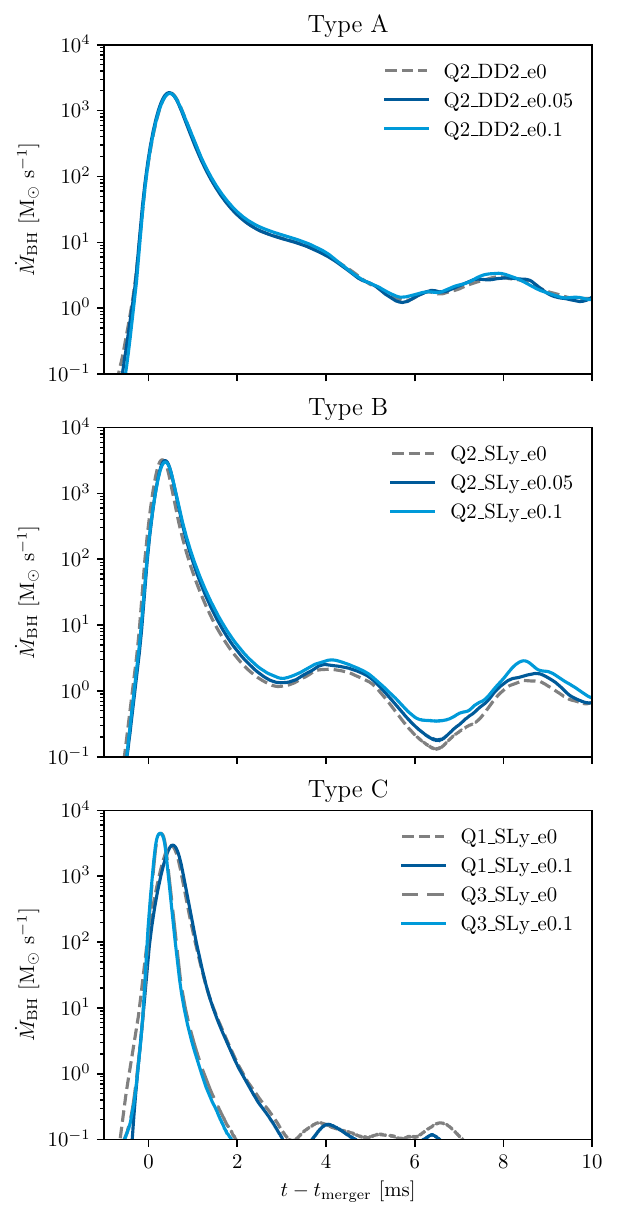}
    \caption{Rate of accretion onto the BH during merger and after merger, categorized by the merger type.}
    \label{fig:accretion}
\end{figure}

In Fig.~\ref{fig:accretion}, we show the accretion rates onto the remnant BH for the configurations simulated here and the corresponding QC ones from Ref.~\cite{Markin:2026eyc}, and use the same accretion type classification. Namely, Type A mergers have their disk perturbed by the spiral density waves and the fallback material, Type B mergers show prompt disk circularization, and Type C mergers correspond to the NS plunging directly into the BH.

For the Type~A and Type~C systems, the accretion behavior remains largely unchanged. For the Type~B systems, \verb|Q2_SLy|, the accretion rate is consistently higher for higher eccentricities. This can be attributed to the larger mass of the disks, allowing more matter to be accreted.
Neither the frequency nor the phase of the trapped $g$-mode oscillations changes with eccentricity. This is expected as their frequencies for low-mass disks are dominated by the mass and spin of the remnant BH~\cite{Okazaki:1987}, which, as we find above, do not change with varying eccentricity. This further confirms that the oscillations observed in the disk are indeed trapped $g$ modes.
Since both DD2 and SLy $Q=2$ configurations produce remnant BHs with very similar masses and spins, the corresponding $g$-mode frequencies are very similar for both.

\section{Effect of strong magnetic fields}
\label{sec:magnetic_fields}
To complement our previously presented results, we perform additional simulations of a selected system with the inclusion of magnetized matter to assess the impact of strong magnetic fields on the dynamics and the presence of the trapped $g$ modes in the remnant disk. For that, we pick \verb|Q2_SLy_e0.1|, as the trapped $g$ modes there are strongly pronounced and their periodic feeding of the BH dominates the accretion rate, cf.~Fig.~\ref{fig:accretion}.

To obtain strongly magnetized accretion disks, and due to the insufficient resolution to resolve small-scale amplification of the magnetic fields~\cite{Hayashi:2021oxy,Hayashi:2022cdq}, we set the NS initial magnetic field with the maximum value of the magnetic field strength of $10^{17}$~G in purely poloidal and purely toroidal topologies. Note that the magnetic field is added \textit{ad hoc} and was not solved consistently with the spacetime and matter, and, therefore, constraint-violating. However, we rely on constraint-damping properties of Z4c and divergence cleaning to drive the solution to a self-consistent state.

A purely poloidal field is set by assigning the components of the vector potential $A_i$ to
\begin{subequations}
    \begin{align}
    A_x &= -(y-y_\mathrm{NS}) A_\mathrm{pol} \max{(p-p_\mathrm{cutoff}, 0)}^2 \\
    A_y &= (x-x_\mathrm{NS}) A_\mathrm{pol} \max{(p-p_\mathrm{cutoff}, 0)}^2 \\
    A_z &= 0,
\end{align}
\end{subequations}

where $x$ and $y$ are the cell coordinates, $x_\mathrm{NS}$ and $y_\mathrm{NS}$ are the coordinates of the NS center, $A_\mathrm{pol}=94940$ is a normalization constant, $p$ is the pressure at the cell, and $p_\mathrm{cutoff}$ is the cutoff pressure set to $0.004$ of the maximum pressure of the NS in the initial data.

The purely toroidal field is defined in terms of the magnetic field components, following~Ref.~\cite{Sur:2021awe},

\begin{subequations}
\begin{align}
    B_x &= -(y-y_\mathrm{NS}) B_\mathrm{tor} \max{(p-p_\mathrm{cutoff}, 0)} \\
    B_y &= (x-x_\mathrm{NS}) B_\mathrm{tor} \max{(p-p_\mathrm{cutoff}, 0)} \\
    B_z &= 0,
\end{align}
\end{subequations}
where $B_\mathrm{tor}=27.2$ is the normalization constant. The cutoff pressure $p_\mathrm{cutoff}$ is identical to the one for purely poloidal topology.\footnote{Note that setting magnetic field components directly produces initial data with nonzero divergence. However, divergence sharply reduces to a value comparable with the poloidal case, as it is efficiently dampened by the divergence cleaning scheme.}

Purely poloidal and toroidal magnetic field configurations are known to be unstable~\cite{Tayler:1957,Tayler:1973,Markey:1973,Wright:1973,Flowers:1977}, and most of the energy is lost over Alfvén timescales. For the magnetic field strengths as high as in our setups, $10^{17}$~G, the Alfvénic timescale is on the order of milliseconds, complicating the initialization of the NS earlier in the inspiral. Therefore, we add a magnetic field to the NS at $\sim2$~ms before merger ($t=26.4$~ms).

In our simulations with a purely poloidal field, the magnetic field is expelled from the core, and the neutral line moves towards the crust, settling at the core-crust boundary. While it is qualitatively consistent with Ref.~\cite{Glampedakis:2015fja}, where it was shown that the closed-line poloidal field can be expelled from the core to the crust due to the crust-core spin lag, it is unlikely to be the case, as the NSs in our simulations are nonspinning. However, the piecewise-polytropic representation of the EOS has discontinuities in the adiabatic index at the dividing densities and thus discontinuities in the speed of sound. While it is generally of less relevance for the hydrodynamical evolution, it can alter the evolution of magnetic fields: the discontinuities in the speed of sound and thus the restoring forces can lead to perturbations in the velocity field at this boundary. Since the magnetic field flux is frozen into the fluid, this leads to amplification of the magnetic field due to the dynamo effect (stretch-twist-fold mechanism). Therefore, such magnetic field amplification is unphysical as it stems from the way the EOS was constructed and depends on the choice of the dividing density.

\subsection{Disk dynamics}
\begin{figure}[t]
    \centering
    \includegraphics[width=1\linewidth]{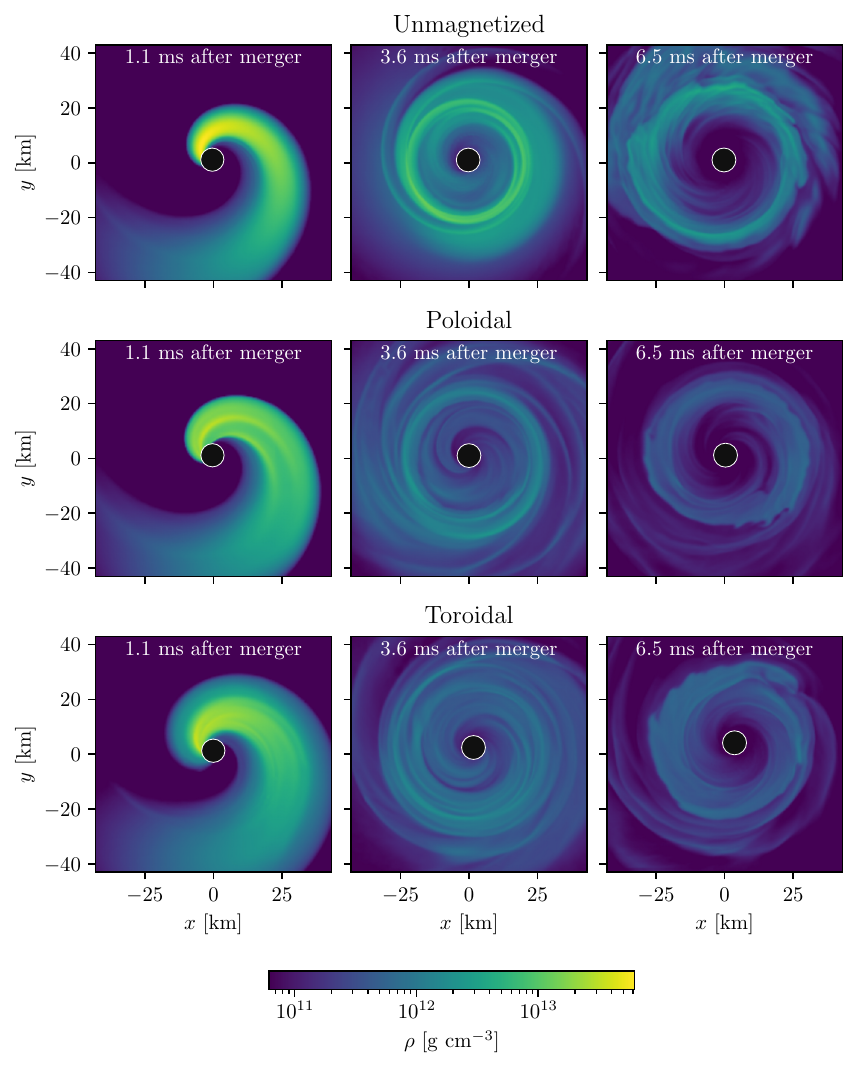}
    \caption{Matter density in the equatorial plane for different stages of merger and disk evolution for unmagnetized, magnetized with the initial NS magnetic field topology being either poloidal or toroidal. The black circle with a white boundary represents the BH horizon.}
    \label{fig:disk_over_time}
\end{figure}

First, we describe qualitative changes that strong magnetic fields introduce into the disk dynamics. In Fig.~\ref{fig:disk_over_time}, we show density slices in the equatorial plane for the unmagnetized and for both magnetized cases. Already at $1.1$~ms after merger, there are noticeable changes in the spiral arm structure: in the magnetized cases, it is less dense and more spatially extended due to the contributions of magnetic pressure. In the toroidal case, the impact is stronger than for the poloidal case: the spiral arm is less dense and has a low-density outer cocoon. This can be explained by the magnetic energy of the toroidal setup being primarily concentrated in the NS crust and the flux-freezing condition of ideal GRMHD. These conditions leave the low-density material from the crust more highly magnetized and having larger magnetic pressure than its core counterpart.

Overall, after their formation, the disks for the magnetized cases are less dense and have less pronounced structure than in the unmagnetized case. Yet, the magnetized disks have visually very similar morphologies and densities.

\begin{figure}[t]
    \centering
    \includegraphics[width=1\linewidth]{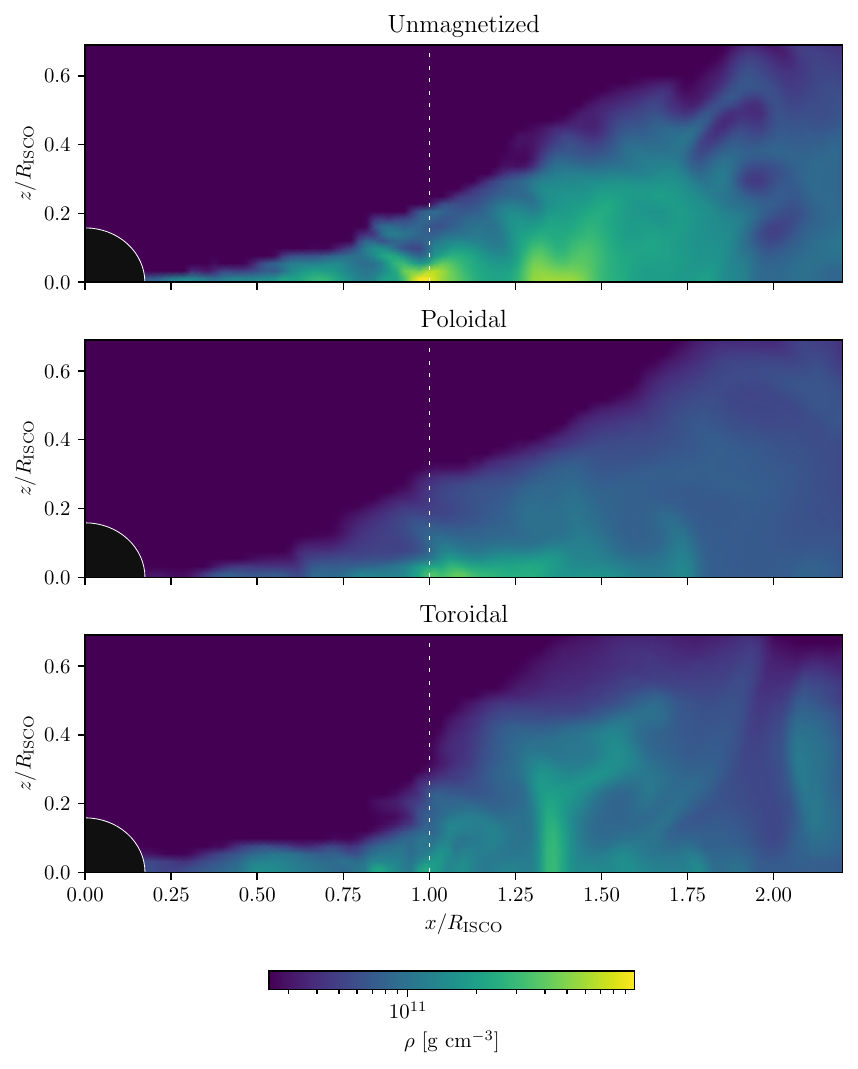}
    \caption{Matter density of the disk in the $x$-$z$ plane at $10.18$~ms after merger for unmagnetized, magnetized with the initial NS magnetic field topology being either poloidal or toroidal. The spatial dimensions are in units of $R_\mathrm{ISCO}$, the radius of the innermost stable circular orbit of the remnant BH (vertical dashed lines). The black circle with a white boundary represents the BH horizon.}
    \label{fig:disk_xz}
\end{figure}

In Fig.~\ref{fig:disk_xz}, we show the vertical ($x$-$z$) slices through the disks at $10.18$~ms after merger. It is evident that the magnetized disks are less dense and have a more dilute structure than the unmagnetized disk, also vertically. Despite the overall lower densities, the disk masses reported in~Tab.~\ref{table:remnant_properties} are higher for magnetized disks. Because we define the disk as the bound matter outside the BH horizon, its mass accounts for the low-density fallback material, which extends vertically due to strong magnetic pressure. 

Analysing the remnant properties, we find that strong magnetic fields are capable of boosting matter ejection due to the addition of magnetic pressure. Meanwhile, most of the currently available fitting formulae predicting dynamical ejecta mass and baryonic remnant mass are calibrated to pure hydrodynamical, unmagnetized simulations. Therefore, if the magnetic field is significantly amplified in such BHNS mergers in reality, it would lead to additional unbound matter, making these fitting formulae potentially underpredicting ejecta mass. We stress, however, that realistic NSs would have significantly weaker surface magnetic fields ($10^8-10^{12}$~G) during inspiral~\cite{Lorimer:2008se,Tauris:2017omb,Kruckow:2018slo}, and likely would be dynamically unimportant to alter mass ejection~\cite{Giacomazzo:2009mp,Chawla:2010sw}.

\subsection{Magnetic field properties}
For the BHNS remnant accretion disks to be strongly magnetized, the magnetic fields must be amplified from $10^8-10^{12}$~G~\cite{Lorimer:2008se,Tauris:2017omb,Kruckow:2018slo} during and after the merger. The primary driver for such an amplification is believed to be magneto-rotational instability (MRI)~\cite{Balbus:1991,Balbus:1998ja}. However, many GRMHD simulations of BHNS mergers are not capable of resolving MRI during merger due to insufficient resolution~\cite{Chawla:2010sw,Etienne:2011ea,Etienne:2012te,Kiuchi:2015qua,Ruiz:2018wah,Hayashi:2021oxy,Hayashi:2022cdq}.

\begin{figure}[htp]
    \centering
    \includegraphics[width=1\linewidth]{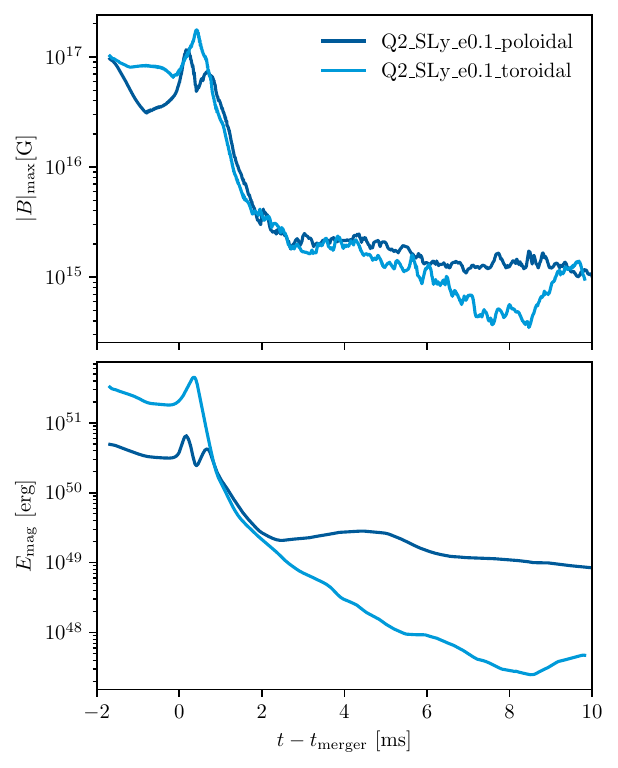}
    \caption{Top: maximum magnetic field strength $|B|_\mathrm{max}$ over time for both magnetized configurations. Bottom: the total magnetic energy $E_\mathrm{mag}$ over time for both magnetized configurations.}
    \label{fig:Bmax_Emag}
\end{figure}

In Fig.~\ref{fig:Bmax_Emag}, we show the evolution of the maximum magnetic field strength and total magnetic energy since the introduction of the magnetic field until $10$~ms after merger.
First, both the poloidal and toroidal configurations start rapidly losing their maximum magnetic field strength due to their inherent instabilities. Note that due to a different topology, the toroidal field stores almost ten times more energy than the poloidal one with the same maximum field strength.

Importantly, the initial increase in the maximum magnetic field strength occurring before the merger is spurious and related to the trapping and artificial amplification of magnetic fields at the crust-core boundary, due to the discontinuity in the sound speed mentioned above. Nevertheless, here, we assume that it has only a minor impact on the dynamics of the magnetic field at and after the merger.

During the merger, the magnetic field is amplified in both cases and reaches values $\sim 10^{17}~\mathrm{G}$. After that, most of the NS matter, primarily located in the core, falls into the BH together with the magnetic fields frozen into it. Therefore, the maximum magnetic field strength falls almost two orders of magnitude down to $\sim 10^{15}~\mathrm{G}$ for both configurations. This behavior is consistent with the one found in Ref.~\cite{Etienne:2011ea} for BHNS mergers with strongly magnetized NSs.

Here, we assess whether MRI is resolved in our simulations by comparing the wavelength of the fastest-growing axisymmetric MRI mode $\lambda^z_\mathrm{MRI}$ to the grid spacing $\Delta x$. We do so by calculating the corresponding quality factor~\cite{Noble:2010mm,Hayashi:2022cdq},
\begin{equation}
    Q^z_\mathrm{MRI}= \frac{ \lambda^z_\mathrm{MRI} }{\Delta x} = v_A^{z} \frac{2 \pi}{\Omega} \frac{1}{\Delta x} = \frac{b^z}{\sqrt{\rho h + b_a b^a}} \frac{2 \pi}{\Omega} \frac{1}{\Delta x},
\end{equation}
where $v_A^z$ is the polar component of the local Alfvén velocity, $\rho$ is the rest-mass density, $\Omega$ is the angular frequency, and $b^z$ is the polar component of the magnetic field strength in the fluid frame.

MRI is to be considered well-resolved if the fastest-growing MRI mode is consistently covered by at least $10$ grid points, i.e., $Q^z_\mathrm{MRI}\sim10$~\cite{Sano:2003bf}. For the accretion disks formed in our simulations, the quality factor remains below $Q^z_\mathrm{MRI}\sim 5$, i.e., the MRI is not well resolved. Resolving MRI would require a substantial increase in resolution\footnote{For instance, attaining the quality factor of $Q^z_\mathrm{MRI}\sim 20$, confidently above the threshold, would require $256$ times more expensive simulations.}, and remains a long-standing bottleneck of GRMHD simulations.
On the other hand, further increasing the \textit{ad hoc} initial magnetic field for the NS would lead to violent readjustment of the stellar structure and change the inspiral dynamics itself~\cite{Giacomazzo:2009mp}. Even in our case of a toroidal initial field, where most of the magnetic energy is concentrated closer to the NS surface, it leads to additional matter ejection and significant changes to the merger dynamics, cf.~Fig.~\ref{fig:disk_over_time}, and produces a large kick to the remnant BH.

\subsection{Effects on trapped $g$-mode oscillations}

\begin{figure}[htp]
    \centering
    \includegraphics[width=1\linewidth]{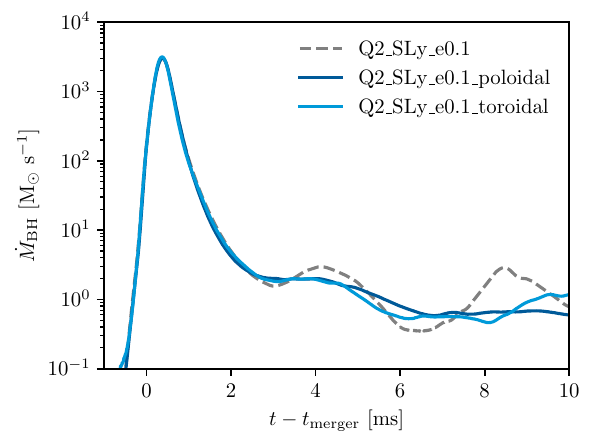}
    \caption{
    Rate of accretion onto the BH during merger and after merger for the Q2\_SLy\_e0.1 binary for the unmagnetized case, and magnetized cases initialized with purely poloidal and purely toroidal magnetic fields for the NS before merger.}
    \label{fig:accretion_magnetized}
\end{figure}

In our previous study~\cite{Markin:2026eyc} and in Sec.~\ref{section:accretion}, we have identified trapped $g$-mode oscillations that modulate the accretion rate by periodically pushing matter onto the BH. These oscillations are analytically predicted for unmagnetized accretion disks~\cite{Okazaki:1987,1992ApJ...393..697N,Perez:1996ti,Wagoner:1998hh,Abramowicz:2011xu,Brink:2015roa}. However, both analytical and numerical studies~\cite{Fu:2008iw,Reynolds:2008wp,Ortega-Rodriguez:2015dpa} predicted that the trapped $g$ modes might be destroyed for strongly magnetized disks. Here, we verify these predictions using our simulations with highly magnetized disks.

In Fig.~\ref{fig:accretion_magnetized}, we show the accretion rates onto the BH for the cases with no magnetic field, with purely poloidal, and purely toroidal magnetic field configurations for the NS. While in the unmagnetized case the accretion rate modulations are well-pronounced and are directly connected with the trapped $g$-mode oscillations in the disk (cf.~Sec.~\ref{section:accretion}), in both cases with a strongly magnetized disk these oscillations are efficiently damped. However, some disk oscillations might still be marginally present, as there is a minor oscillatory imprint onto the accretion rate with a frequency similar to that in the unmagnetized case. We therefore conclude that if the accretion disks of such BHNS mergers indeed have their magnetic field amplified to high values of $|B|\approx10^{15}\mathrm{G}$, it is unlikely that such modulations will be detectable observationally. Conversely, the presence of quasi-periodic oscillations in the observables of confirmed BHNS mergers, such as prompt emission of short gamma-ray bursts, would indicate that the accretion disks might not be sufficiently magnetized.

\subsection{Disk magnetization and polar funnel formation}
\begin{figure}[!htp]
    \centering
    \includegraphics[width=1\linewidth]{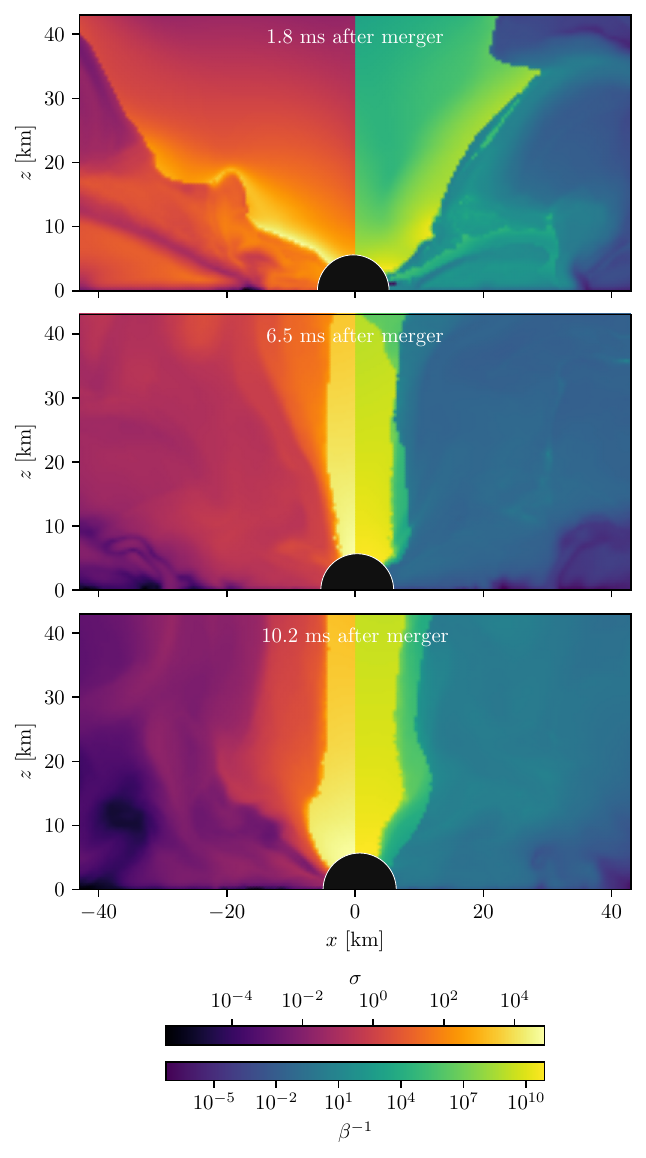}
    \caption{\textit{Left}: Magnetization $\sigma$ of the matter on a slice along the $x$-$z$ coordinate direction. \textit{Right}: Inverse plasma beta $\beta^{-1}$ of the matter on a slice along the $x$-$z$ coordinate direction. The black circle represents the apparent horizon of the BH.}
    \label{fig:magnetization}
\end{figure}

In Fig.~\ref{fig:magnetization}, we show the magnetization parameter $\sigma = b^2/{2\rho}$ and inverse plasma beta $\beta^{-1}=p_\mathrm{mag}/p_\mathrm{gas}$ of the accretion disk in the $x$-$z$ (polar) plane. We note that both parameters, although different, display highly similar structure; therefore, here, we discuss only magnetization $\sigma$. 
In the equatorial plane, the magnetization is low, $\sigma < 1$, i.e., below the equipartition state. In the polar direction and within a narrow funnel, on the other hand, the magnetization can reach extreme values of $\sigma \sim 3 \times 10^{4}$. Because of the absence of mechanisms to deposit matter into polar direction (neutrino winds are one such mechanism, but we do not include neutrino radiation here), the baryonic density there is very low, $\rho \sim 10^{3} ~\mathrm{g/cm^{3}}$. However, these density values are close to the artificial atmosphere density floors, $\sim 620~\mathrm{g/cm^{3}}$, and therefore can be used only as an indicator of a lower bound of magnetization. Hence, the presented magnetization is likely underestimated.
Still, it is sufficiently high to provide a favourable environment for launching a highly relativistic jet~\cite{Blandford:1977ds,Narayan:2003by,Tchekhovskoy:2011zx}, thought to be the engine of short gamma-ray bursts (sGRBs)~\cite{Paschalidis:2014qra,Hayashi:2022cdq,Gottlieb:2023est}.

Additionally, we calculate the dimensionless magnetic flux threading the apparent horizon (also known as the MADness parameter)~\cite{Tchekhovskoy:2011zx}, defined as
\begin{equation}
    \phi_{\rm BH} = \frac{\Phi_{\rm BH}}{\sqrt{\dot{M} r_g^2 c }} \approx \frac{\Phi_\mathrm{BH}}{\sqrt{\dot{M}_\textrm{BH}} M_\textrm{BH}},
\end{equation}

where
\begin{equation}
    \Phi_\mathrm{BH} = \frac{1}{2} \int_\mathrm{AH} |B^r| dA
\end{equation}
is the magnetic flux threading the apparent horizon.
For both of our magnetized simulations, $\phi_\mathrm{BH}$ remains low and never exceeds $2.5$. Therefore, the accretion disks remain well within the standard and normal evolution (SANE) mode, where the accretion is matter-dominated. This is further confirmed by the low values of inverse plasma beta, being $\beta^{-1}<1$ throughout the disk. This is primarily due to the relatively weak magnetic fields in the disk, which in reality would likely be noticeably higher due to the amplification that cannot be captured in our simulations because of insufficient resolution.

\section{Conclusion}
\label{section:conclusion}
We have simulated a set of BHNS mergers with mass ratios $Q \in \{1,2,3\}$ with orbital eccentricities $e_\mathrm{GW}^\mathrm{20Hz} \in [0.64, 0.81]$. Such eccentricities are around five times higher than the ones inferred for the BHNS merger, GW200105. These larger eccentricities were chosen to explore the region of parameter space in which eccentricity effects remain prominent in the late inspiral and merger despite efficient GW-driven circularization in the early inspiral.

Comparing with waveform models, we find that the SEOBNRv5EHM model can describe the inspiral of such eccentric BHNS systems as long as the tidal contributions are negligible. This is the case for most of the simulations up to the last six GW cycles, and throughout the entire waveform for the case where the NS plunges into the BH as a whole, resembling a BBH merger signal. For the inspiral-only post-Newtonian pyEFPEHM model, the dephasing oscillates in the late inspiral outside the NR error band, with these oscillations correlating with the radial period of the binary, therefore suggesting that some eccentric contributions to the phase remain unaccounted for in the model. The inclusion of QC tidal corrections to the model improves the agreement with the NR data, especially for the cases with larger effective tidal deformability. Incorporation of eccentric tidal contributions to pyEFPEHM has a small but noticeable impact on the waveform in cases with larger effective tidal deformability, and a negligible impact in the others. It is likely that for systems where the NS has larger tidal deformability and/or higher eccentricity, eccentric tidal corrections would become more important for improving agreement with the numerical waveforms.
Finally, we find that pyEFPE does not show good agreement with the numerical waveforms and has large dephasing ($\gtrsim 1$~rad) shortly after the alignment window, which primarily stems from the employment of lower post-Newtonian orders for the phasing contributions in the model.

Analyzing the remnant properties, we find that for the stiff EOS, the mass of dynamical ejecta and accretion disk can become both larger or smaller with higher eccentricity. We suggest that these changes are likely to depend on the value of the phase of the radial period (true anomaly) at the time of tidal disruption and matter ejection. Further studies specifically targeting the effect of varying true anomaly would be necessary to confirm this hypothesis.

We identify global trapped $g$-mode oscillations in the disk, consistent with our observations in the QC cases~\cite{Markin:2026eyc}. These oscillations periodically push matter onto the remnant BH, modulating its accretion rate. The frequencies of trapped $g$ modes are expected to be set by the mass and spin of the remnant BH, regardless of the disk mass and composition. The frequencies of these trapped $g$-mode oscillations in the eccentric cases match closely with the ones in the QC cases. These frequencies do not change with eccentricity, matching the observation of unaffected remnant BH properties, and, therefore, further confirming the expectations of trapped $g$ modes. If these oscillations are present and capable of modulating the observables, such as gamma-ray burst prompt emission, they would serve as a clear, eccentricity-independent probe for the remnant BH properties and an indicator of the existence of BHNS systems with mass ratios close to unity, where such massive accretion disks may form. 

Additionally, we perform simulations with magnetized NS matter in poloidal and toroidal topologies, and find that when disks are strongly magnetized ($10^{15}$~G), the trapped $g$ modes are successfully damped, e.g., \cite{Fu:2008iw,Reynolds:2008wp,Ortega-Rodriguez:2015dpa}. As the postmerger accretion disks are expected to be strongly magnetized, these oscillations would be unlikely to be detected. Conversely, an observation of such oscillations would indicate weakly magnetized disks shortly after merger. In that case, one would be able to constrain the magnetic field amplification mechanisms or the initial magnetic field of the NS in such a system. 

\section*{Data Availability}
The data that support the findings of this article are openly available~\cite{markin_2026_23366197,markin_2026_23368052}.
The repository containing the waveform model, with eccentric tidal corrections and quasi-universal relations, is available at Ref.~\cite{pyEFPEHM_repo}.

\section*{Acknowledgements}
\label{section:acknowledgements}
I.~M. is grateful to Jakob Stegmann for the insights on formation channels of eccentric compact binaries, and to Guillermo Lara and Maxence Corman for helpful comments regarding waveform alignment.
I.~M. and T.~D. gratefully acknowledge support from the Deutsche Forschungsgemeinschaft (DFG) under the project 504148597 (DI 2553/7). T.~D. acknowledges funding from the European Union (ERC, SMArt, 101076369).
GM's research is supported in part by the European Research Council (ERC) Horizon Synergy Grant “Making Sense of the Unexpected in the Gravitational-Wave Sky” grant agreement no. GWSky–101167314.
Views and opinions expressed are those of the authors only and do not necessarily reflect those of the European Union or the European Research Council. Neither the European Union nor the granting authority can be held responsible for them. 
The authors gratefully acknowledge the computing time granted by the Resource Allocation Board and provided on the supercomputer Emmy at NHR-Nord@Göttingen as part of the NHR infrastructure. Most calculations for this research were conducted with computing resources under the project bbp00049.
\appendix

\section{Alignment parameters}
\label{sec:alignment_values}

\begin{table*}[t]
    \centering
    \begin{tabular}{l|c|c|c|c|c|c}
Name & $e_\mathrm{GW}^\mathrm{20Hz}$ & SEOBNRv5EHM & pyEFPE & pyEFPEHM & pyEFPEHM (QC tides) & pyEFPEHM (ecc. tides) \\
\hline
\verb|Q1_SLy_e0.1| & $0.81$ & $0.81$ & $0.84$ & $0.80$ & $0.80$ & $0.80$ \\
\verb|Q2_SLy_e0.05| & $0.64$ & $0.64$ & $0.70$ & $0.63$ & $0.63$ & $0.62$ \\
\verb|Q2_SLy_e0.1| & $0.77$ & $0.77$ & $0.81$ & $0.76$ & $0.76$ & $0.76$ \\
\verb|Q3_SLy_e0.1| & $0.73$ & $0.73$ & $0.78$ & $0.72$ & $0.72$ & $0.72$ \\
\verb|Q2_DD2_e0.05| & $0.64$ & $0.64$ & $0.71$ & $0.63$ & $0.63$ & $0.63$ \\
\verb|Q2_DD2_e0.1| & $0.77$ & $0.77$ & $0.81$ & $0.76$ & $0.76$ & $0.76$ \\
    \end{tabular}
    \caption{The optimal waveform model eccentricity parameters $e_0$ at $20$~Hz found during the waveform alignment. The first column shows the GW-eccentricity $e_\mathrm{GW}^\mathrm{20Hz}$ from Tab.~\ref{table:configurations} to enable direct comparison.}
    \label{tab:alignment_eccentricities}
\end{table*}
In Tab.~\ref{tab:alignment_eccentricities}, we list the eccentricity parameters at 20~Hz, $e_0^\mathrm{20Hz}$, of the waveform models after alignment. The eccentricity parameters of SEOBNRv5EHM exactly match the GW-eccentricity $e_\mathrm{GW}^\mathrm{20Hz}$ for all cases. The eccentricities are very close between the SEOBNRv5EHM and pyEFPEHM models, while noticeably different for pyEFPE, which is a direct consequence of a poorer alignment.

\section{Convergence}
\label{section:convergence}
\subsection{Gravitational waveforms}

\begin{figure*}[!htp]
    \centering
    \includegraphics[width=1\linewidth]{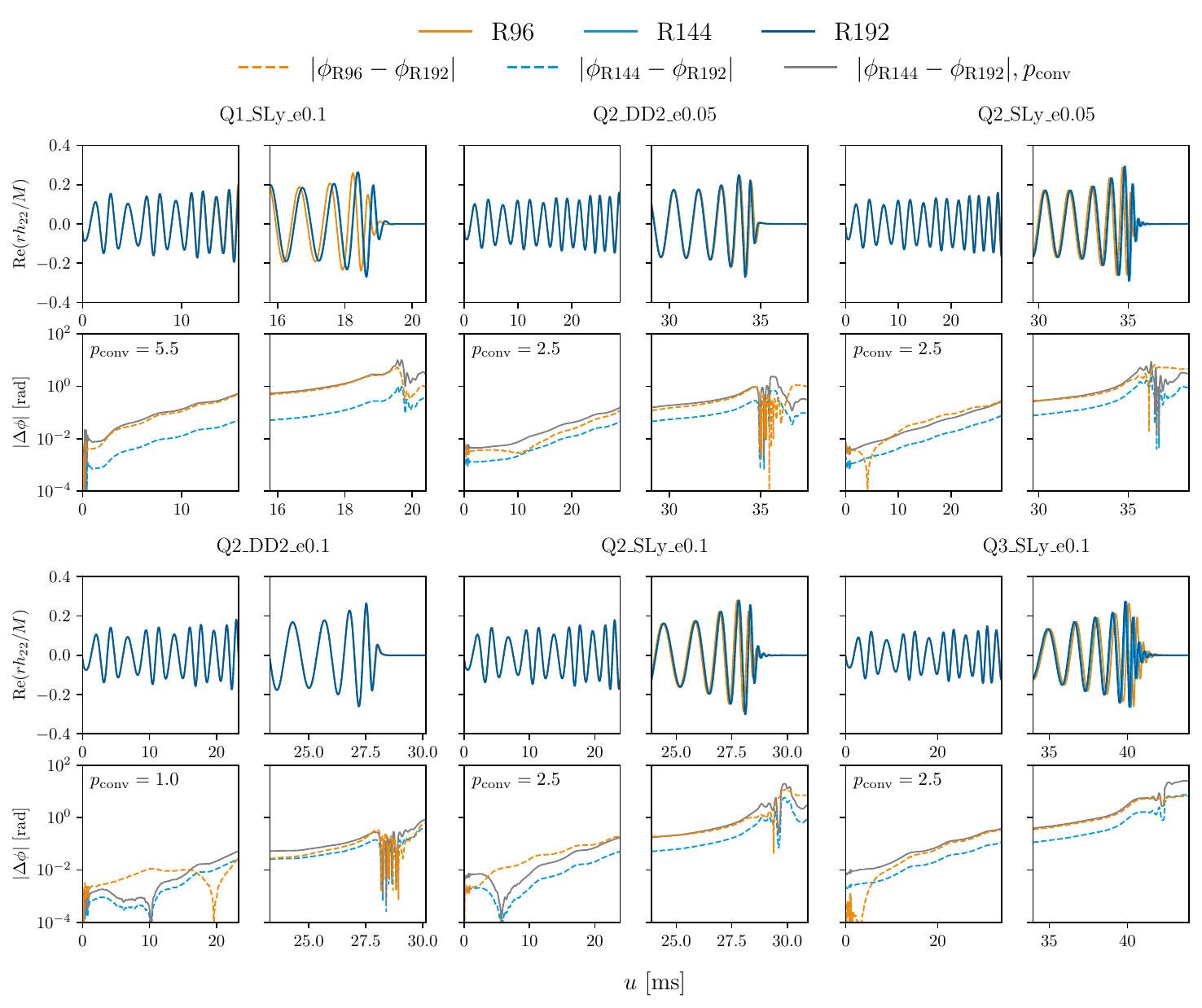}
    \caption{Top panels: The NR waveforms for different resolutions. Bottom panels: Dephasing of the waveforms at low and medium resolution relative to the one at high resolution, and the rescaled dephasing between waveforms at the medium and high resolutions assuming a convergence order $p_\mathrm{conv}$.}
    \label{fig:nr_waveforms_convergence}
\end{figure*}

In Fig.~\ref{fig:nr_waveforms_convergence}, we plot the dephasing between the numerical waveforms at low (R96), medium (R144), and high (R192) resolutions. As in our previous study~\cite{Markin:2026eyc}, most configurations show an apparent convergence order of $p_\mathrm{conv}=2.5$, and follow this order relatively cleanly throughout the inspiral and merger. Nevertheless, there are two outliers: \verb|Q1_SLy_e0.1| and \verb|Q2_DD2_e0.1|. 
For \verb|Q1_SLy_e0.1|, the apparent convergence order reaches $p_\mathrm{conv}=5.5$, mirroring the QC \verb|SLy_mNS1.4_mBH1.4| of Ref.~\cite{Markin:2026eyc}, where it had $p_\mathrm{conv}=6$. Interestingly, the convergence of the waveform for this case follows the apparent convergence order remarkably well.
For \verb|Q2_DD2_e0.1|, the phase difference between different resolutions is small, $\lesssim0.1$~rad, throughout inspiral. This is in contrast with the corresponding \verb|DD2_mNS1.4_mBH2.8| case in Ref.~\cite{Markin:2026eyc}, which followed the apparent convergence order $p_\mathrm{conv}=2.5$. While the exact reason for the coinciding waveforms is unknown, this might suggest that this case has already reached a converged state. However, a higher-resolution study would be required to confirm this.

\subsection{Remnant properties}
\label{section:convergence_remnant_properties}

In Tab.~\ref{table:big_convergence_table}, we provide the values for the baryonic remnant mass (disk and dynamical ejecta), and remnant BH properties for all three resolutions. All the quantities form convergent series. The baryonic remnant mass decreases with the resolution in all cases, except the equal-mass-ratio system, where the opposite is true.
The mass of the remnant BH always increases with the resolution, while its spin remains largely unchanged.

\begin{table*}[!htp]
    \centering
    \begin{tabular}{l|c|c|c|ccc|ccc}
       Name  & $M_{\mathrm{b,rem}}^{\mathrm{R96}}$ & $M_{\mathrm{b,rem}}^{\mathrm{R144}}$ & $M_{\mathrm{b,rem}}^{\mathrm{R192}}$ & $M_{\mathrm{BH,rem}}^{\mathrm{R96}}$ & $M_{\mathrm{BH,rem}}^{\mathrm{R144}}$ & $M_{\mathrm{BH,rem}}^{\mathrm{R192}}$ &
       $\chi_{\mathrm{BH,rem}}^{\mathrm{R96}}$ & $\chi_{\mathrm{BH,rem}}^{\mathrm{R144}}$ & $\chi_{\mathrm{BH,rem}}^{\mathrm{R192}}$ \\
       \hline
\verb|Q1_SLy_e0.1| & $1.52 \times 10^{-3}$ & $1.60 \times 10^{-3}$ & $1.87 \times 10^{-3}$ & $2.725$ & $2.734$ & $2.736$ & $0.812$ & $0.814$ & $0.814$ \\
\verb|Q2_SLy_e0.05 | & $2.15 \times 10^{-2}$ & $1.84 \times 10^{-2}$ & $1.71 \times 10^{-2}$ & $4.064$ & $4.073$ & $4.075$ & $0.670$ & $0.670$ & $0.669$ \\
\verb|Q2_SLy_e0.1 | & $2.93 \times 10^{-2}$ & $2.64 \times 10^{-2}$ & $2.51 \times 10^{-2}$ & $4.059$ & $4.067$ & $4.069$ & $0.670$ & $0.669$ & $0.669$ \\
\verb|Q2_DD2_e0.05 | & $8.80 \times 10^{-2}$ & $8.45 \times 10^{-2}$ & $8.25 \times 10^{-2}$ & $4.031$ & $4.035$ &  $4.037$ & $0.668$ & $0.669$ & $0.670$ \\
\verb|Q2_DD2_e0.1 | & $9.25 \times 10^{-2}$ & $8.93 \times 10^{-2}$ & $8.82 \times 10^{-2}$ & $4.028$ & $4.032$ & $4.033$ & $0.669$ & $0.670$ &  $0.670$ \\
\verb|Q3_SLy_e0.1 | & $2.99 \times 10^{-3}$ & $9.26 \times 10^{-4}$
 & $4.03 \times 10^{-4}$ & $5.452$ & $5.452$ & $5.452$ & $0.550$ & $0.549$ & $0.549$ \\

    \end{tabular}
    \caption{Baryonic mass of the remnant outside the BH horizon $M_{\mathrm{b,rem}}$, remnant BH mass $M_{\mathrm{BH,rem}}$ and spin $\chi_{\mathrm{BH,rem}}$ at three different resolutions extracted at $10$~ms after merger.} 
    \label{table:big_convergence_table}
\end{table*}

\section{Eccentric tidal corrections}
\label{sec:appendix:ecc_tides}

The pyEFPEHM model~\cite{Morras:2026fho} incorporates adiabatic tidal effects only in the QC limit, using the 7.5PN results for zero component spins of Ref.~\cite{Dones:2024odv} and the 6.5PN spin-tidal ones of Refs.~\cite{Abdelsalhin:2018reg,Castro:2022mpw}. Here we describe the two additions made for this work. First, we replace the QC adiabatic tidal coefficients by their eccentricity-dependent counterparts, which were recently derived through relative 2.5PN order in Refs.~\cite{Henry:2025uta,Henry:2026bqh}. Furthermore, the matter parameters that are not taken directly from the EOS are obtained from the quadrupolar tidal deformability through quasi-universal relations, as described in Appendix~\ref{sec:appendix:QUR}.

We follow the notation of Refs.~\cite{Morras:2025nlp,Morras:2026fho}. The binary is described by the symmetric mass ratio $\nu = m_1 m_2/M^2$ with $m_1$, $m_2$ being the mass of the primary and the secondary object, respectively, $M = m_1 + m_2$, the harmonic-coordinate time eccentricity $e$, and the PN parameter
\begin{equation}
    y = \frac{(M \omega)^{1/3}}{\sqrt{1 - e^2}} \, ,
    \label{eq:ydef_tides}
\end{equation}
\noindent with $\omega$ the mean orbital angular velocity. Note that $e$ is the time eccentricity denoted $e_t$ in Refs.~\cite{Henry:2025uta,Henry:2026bqh}, and differs in general from the eccentricity estimators quoted in Table~\ref{table:configurations}. Introducing the rescaled time derivative $\D = M (1-e^2)^{-3/2}\, \mathrm{d}/\mathrm{d}t$, the secular evolution equations and the periastron advance $k$ are written as the PN series~\cite{Morras:2025nlp}
\begin{subequations}
\label{eq:RR_eqs_tides}
\begin{align}
    \D y   &= \nu y^9 \left(a_0 + \sum_{n \geq 2} a_n y^n \right) \, , \label{eq:RR_eqs_tides:y} \\
    \D e^2 &= - \nu y^8 \left(b_0 + \sum_{n \geq 2} b_n y^n \right) \, , \label{eq:RR_eqs_tides:e2} \\
    k      &= y^2 \sum_{n \geq 0} k_n y^n \, . \label{eq:RR_eqs_tides:k}
\end{align}
\end{subequations}

Adiabatic tidal effects first enter the phasing at 5PN order, contributing $a_{n}^\mathrm{T}$ and $b_{n}^\mathrm{T}$ for $n \geq 10$ and $k_{n}^\mathrm{T}$ for $n \geq 8$. We obtain these coefficients from the ancillary material of Refs.~\cite{Henry:2025uta,Henry:2026bqh}: the secular evolution of the orbital elements $\langle \dot{x}\rangle$ and $\langle \dot{e}_t \rangle$ from Ref.~\cite{Henry:2026bqh}, and the periastron advance $K = 1 + k$ from Ref.~\cite{Henry:2025uta}, both in terms of the quasi-Keplerian variables $(x, e_t)$ with $x = (M\omega)^{2/3} = (1-e^2) y^2$. The change of variables to $(y, e^2)$ follows from Eq.~\eqref{eq:ydef_tides},
\begin{subequations}
\label{eq:Henry_to_EFPE}
\begin{align}
    \D y   &= M \, \frac{\dot{x} + 2 y^2 e \dot{e}}{2 y \left(1-e^2\right)^{5/2}} \, , \label{eq:Henry_to_EFPE:y} \\
    \D e^2 &= M \, \frac{2 e \dot{e}}{\left(1-e^2\right)^{3/2}} \, . \label{eq:Henry_to_EFPE:e2}
\end{align}
\end{subequations}

\noindent The tidal parameters of Refs.~\cite{Henry:2025uta,Henry:2026bqh}, denoted there by $\widetilde{\mu}^{(\ell)}_\pm$ and $\widetilde{\sigma}^{(\ell)}_\pm$, are related to the mass-weighted combinations of Ref.~\cite{Morras:2026fho} by
\begin{align}
    \widetilde{\mu}^{(2)}_{+} =& \, \frac{\Lambda^{(2)}_{A,\mathrm{av}}}{32} \, , &
    \delta\mu \, \widetilde{\mu}^{(2)}_{-} =& \, \frac{2 \nu \Lambda^{(2)}_{B,\mathrm{av}} - \left(1 - 2 \nu\right) \Lambda^{(2)}_{A,\mathrm{av}}}{32} \, , \nonumber\\
    \widetilde{\sigma}^{(2)}_{+} =& \, \frac{\Sigma^{(2)}_{A,\mathrm{av}}}{32} \, , &
    \delta\mu \, \widetilde{\sigma}^{(2)}_{-} =& \, \frac{2 \nu \Sigma^{(2)}_{B,\mathrm{av}} - \left(1 - 2 \nu\right) \Sigma^{(2)}_{A,\mathrm{av}}}{32} \, , \nonumber\\
    \widetilde{\mu}^{(3)}_{+} =& \, \frac{\Lambda^{(3)}_{A,\mathrm{av}}}{128} \, ,
    \label{eq:tidal_param_map}
\end{align}

\noindent where $\delta\mu = (m_1 - m_2)/M$. Following Ref.~\cite{Morras:2026fho}, we define for a generic dimensionless tidal deformability $K^{(\ell)}_i$ of body $i$
\begin{subequations}
\label{eq:Kcombs}
\begin{align}
    K^{(\ell)}_{A,\mathrm{av}} =& \, 2^{2 \ell} \frac{m_1^{2 \ell} m_2 K^{(\ell)}_1 + m_2^{2 \ell} m_1 K^{(\ell)}_2}{(m_1 + m_2)^{2 \ell + 1}}  \, , \\
    K^{(\ell)}_{B,\mathrm{av}} =& \, 2^{2 \ell} \frac{m_1^{2 \ell + 1} K^{(\ell)}_1 + m_2^{2 \ell + 1} K^{(\ell)}_2}{(m_1 + m_2)^{2 \ell + 1}}  \, ,
\end{align}
\end{subequations}

\noindent and $\Lambda^{(2)}_i$, $\Lambda^{(3)}_i$, and $\Sigma^{(2)}_i$ are the mass-quadrupolar, mass-octupolar, and current-quadrupolar dimensionless tidal deformabilities, given in terms of the tidal Love numbers $k^{(\ell)}_i$ and $j^{(\ell)}_i$ by~\cite{Hinderer:2007mb,Henry:2020ski}
\begin{subequations}
\label{eq:Lovedefs}
\begin{align}
    \Lambda^{(\ell)}_i &= \frac{2}{(2 \ell - 1)!!} k_i^{(\ell)} \left( \frac{c^2 R_i}{G m_i} \right)^{2 \ell + 1} \, , \\
    \Sigma^{(\ell)}_i  &= \frac{\ell - 1}{4 (\ell+2) (2 \ell - 1)!!} j_i^{(\ell)} \left( \frac{c^2 R_i}{G m_i} \right)^{2 \ell + 1} \, ,
\end{align}
\end{subequations}

\noindent with $R_i$ the radius of body $i$. All deformabilities vanish for black holes, and $\Lambda^{(2)}_i$ is the quantity listed as $\Lambda_\mathrm{NS}$ in Table~\ref{table:configurations}.

With these conventions, the eccentric adiabatic tidal contributions to Eq.~\eqref{eq:RR_eqs_tides:y} are

\begin{widetext}
\begin{subequations}
\label{eq:aTecc}
\begin{align}
    a_{10}^\mathrm{T} =& \bigg[\frac{144}{5} + \frac{836}{5} e^2 + \frac{31047}{160} e^4 + \frac{14937}{320} e^6 + \frac{943}{640} e^8\bigg] \Lambda^{(2)}_{A,\mathrm{av}} + \bigg[\frac{12}{5} + \frac{351}{10} e^2 + \frac{549}{8} e^4 + \frac{381}{16} e^6 + \frac{99}{128} e^8\bigg] \Lambda^{(2)}_{B,\mathrm{av}} \nonumber\\
    & + \frac{1 - \sqrt{1 - e^2}}{\sqrt{1 - e^2}} \bigg[-6 + \frac{137}{20} e^2 - \frac{635}{32} e^4 - \frac{431}{320} e^6 + \frac{67}{160} e^8\bigg] \Lambda^{(2)}_{A,\mathrm{av}} \, , \label{eq:aTecc:10} \\
    a_{12}^\mathrm{T} =& \bigg[\bigg(\frac{4421}{140} + \frac{1319937}{560} e^2 + \frac{985073}{128} e^4 + \frac{204209}{40} e^6 + \frac{12368995}{14336} e^8 + \frac{45101}{2560} e^{10}\bigg) - \nu \bigg(\frac{571}{10} + \frac{21935}{12} e^2 \nonumber\\
    & + \frac{1335341}{240} e^4 + \frac{1974391}{480} e^6 + \frac{642493}{960} e^8 + \frac{6329}{384} e^{10}\bigg)\bigg] \Lambda^{(2)}_{A,\mathrm{av}} + \bigg[\bigg(\frac{38}{35} + \frac{90621}{280} e^2 + \frac{217845}{112} e^4 + \frac{231639}{112} e^6 \nonumber\\
    & + \frac{756747}{1792} e^8 + \frac{57501}{7168} e^{10}\bigg) + \nu \bigg(\frac{751}{10} + \frac{885}{8} e^2 - \frac{372337}{320} e^4 - \frac{227927}{128} e^6 - \frac{197483}{512} e^8 - 9 e^{10}\bigg)\bigg] \Lambda^{(2)}_{B,\mathrm{av}} + \bigg[\frac{4148}{15} \nonumber\\
    & + \frac{9978}{5} e^2 + \frac{32397}{10} e^4 + \frac{7791}{8} e^6 + \frac{13431}{64} e^8 + \frac{47}{64} e^{10}\bigg] \Sigma^{(2)}_{A,\mathrm{av}} + \bigg[-\frac{4}{15} - \frac{54}{5} e^2 - \frac{273}{5} e^4 - \frac{119}{2} e^6 - \frac{441}{32} e^8 \nonumber\\
    & - \frac{21}{64} e^{10}\bigg] \Sigma^{(2)}_{B,\mathrm{av}} + \frac{1 - \sqrt{1 - e^2}}{\sqrt{1 - e^2}} \bigg\{ \bigg[-\bigg(\frac{22467}{280} - \frac{6215}{112} e^2 - \frac{28723}{280} e^4 + \frac{1102047}{1120} e^6 - \frac{195451}{17920} e^8 - \frac{57481}{8960} e^{10}\bigg) \nonumber\\
    & + \nu \bigg(\frac{99}{10} - \frac{12529}{120} e^2 + \frac{147}{10} e^4 + \frac{484171}{960} e^6 + \frac{12743}{256} e^8 - \frac{2009}{384} e^{10}\bigg)\bigg] \Lambda^{(2)}_{A,\mathrm{av}} - \nu \bigg(35 - \frac{45}{8} e^2 + \frac{3195}{64} e^4 \nonumber\\
    & + \frac{7075}{128} e^6 - \frac{1175}{64} e^8\bigg) \Lambda^{(2)}_{B,\mathrm{av}} + \bigg[-168 + 15 e^2 - 34 e^4 - \frac{5689}{8} e^6 + \frac{1979}{32} e^8 - \frac{7}{16} e^{10}\bigg] \Sigma^{(2)}_{A,\mathrm{av}} \bigg\} \, , \label{eq:aTecc:12} \\
    a_{13}^\mathrm{T} =& \pi \bigg\{ \frac{576}{5} \theta_{y,\Lambda_A,13} \Lambda^{(2)}_{A,\mathrm{av}} + \frac{48}{5} \theta_{y,\Lambda_B,13} \Lambda^{(2)}_{B,\mathrm{av}} \bigg\} \, , \label{eq:aTecc:13} \\
    a_{14}^\mathrm{T} =& \bigg[\bigg(\frac{6993499}{15120} + \frac{15714269}{1512} e^2 + \frac{1102179223}{15120} e^4 + \frac{52322942411}{483840} e^6 + \frac{3070450891}{86016} e^8 + \frac{86768245}{18432} e^{10} \nonumber\\
    & + \frac{11317927}{172032} e^{12}\bigg) + \nu \bigg(\frac{18677}{105} - \frac{4227115}{672} e^2 - \frac{80707429}{840} e^4 - \frac{4801163293}{26880} e^6 - \frac{8385074287}{107520} e^8 - \frac{768392641}{86016} e^{10} \nonumber\\
    & - \frac{16057529}{107520} e^{12}\bigg) - \nu^2 \bigg(\frac{1239}{10} - \frac{357469}{80} e^2 - \frac{181318513}{3840} e^4 - \frac{392908657}{4608} e^6 - \frac{191582975}{4608} e^8 - \frac{54700813}{11520} e^{10} \nonumber\\
    & - \frac{13669}{144} e^{12}\bigg)\bigg] \Lambda^{(2)}_{A,\mathrm{av}} + \bigg[\bigg(\frac{21011}{630} + \frac{173864}{105} e^2 + \frac{109168823}{6720} e^4 + \frac{158511807}{4480} e^6 + \frac{397260377}{21504} e^8 \nonumber\\
    & + \frac{405549581}{215040} e^{10} + \frac{177965}{14336} e^{12}\bigg) + \nu \bigg(\frac{378769}{560} + \frac{7906847}{672} e^2 + \frac{19089725}{1344} e^4 - \frac{399309413}{10752} e^6 - \frac{1005710919}{28672} e^8 \nonumber\\
    & - \frac{937810319}{215040} e^{10} - \frac{402673}{7168} e^{12}\bigg) - \nu^2 \bigg(\frac{5231}{12} + \frac{4250659}{720} e^2 + \frac{27142819}{5760} e^4 - \frac{256379107}{11520} e^6 - \frac{193185643}{9216} e^8 \nonumber\\
    & - \frac{376791}{128} e^{10} - \frac{1605}{32} e^{12}\bigg)\bigg] \Lambda^{(2)}_{B,\mathrm{av}} + \bigg[60 + \frac{7879}{16} e^2 + \frac{59239}{64} e^4 + \frac{107649}{256} e^6 + \frac{163543}{2048} e^8 + \frac{1015}{512} e^{10}\bigg] \Lambda^{(3)}_{A,\mathrm{av}} \nonumber\\
    & + \bigg[\bigg(\frac{74966}{105} + \frac{4203881}{140} e^2 + \frac{12029685}{112} e^4 + \frac{31471151}{336} e^6 + \frac{12672169}{896} e^8 + \frac{7781601}{2560} e^{10} + \frac{52095}{7168} e^{12}\bigg) \nonumber\\
    & - \nu \bigg(\frac{11854}{15} + \frac{888734}{45} e^2 + \frac{12419029}{180} e^4 + \frac{11093731}{180} e^6 + \frac{3679321}{360} e^8 + \frac{2156525}{1152} e^{10} + \frac{1607}{192} e^{12}\bigg)\bigg] \Sigma^{(2)}_{A,\mathrm{av}} \nonumber\\
    & + \bigg[-\bigg(\frac{299}{315} + \frac{9988}{105} e^2 + \frac{315713}{280} e^4 + \frac{83777}{30} e^6 + \frac{222105}{128} e^8 + \frac{9529}{40} e^{10} + \frac{3317}{1024} e^{12}\bigg) + \nu \bigg(\frac{15506}{15} + \frac{29308}{3} e^2 \nonumber\\
    & + \frac{4011151}{180} e^4 + \frac{10664423}{720} e^6 + \frac{25953361}{5760} e^8 + \frac{1714169}{2880} e^{10} + \frac{973}{192} e^{12}\bigg)\bigg] \Sigma^{(2)}_{B,\mathrm{av}} + \frac{1 - \sqrt{1 - e^2}}{\sqrt{1 - e^2}} \bigg\{ \bigg[\bigg(\frac{5313599}{7560} \nonumber\\
    & + \frac{99753767}{15120} e^2 + \frac{5004769589}{241920} e^4 - \frac{10792769629}{483840} e^6 - \frac{5486137117}{161280} e^8 - \frac{132934883}{80640} e^{10} + \frac{11479}{672} e^{12}\bigg) \nonumber\\
    & - \nu \bigg(\frac{791837}{1680} + \frac{7014449}{1680} e^2 + \frac{5758529}{384} e^4 - \frac{9966857}{672} e^6 - \frac{194204035}{7168} e^8 - \frac{89192267}{71680} e^{10} + \frac{6706709}{107520} e^{12}\bigg) \nonumber\\
    & + \nu^2 \bigg(\frac{3437}{30} + \frac{399151}{480} e^2 + \frac{4898621}{1280} e^4 - \frac{86360387}{23040} e^6 - \frac{36300523}{5760} e^8 - \frac{3070357}{5760} e^{10} + \frac{10193}{288} e^{12}\bigg)\bigg] \Lambda^{(2)}_{A,\mathrm{av}} \nonumber\\
    & + \bigg[\bigg(66 + \frac{7149}{4} e^2 + \frac{78573}{16} e^4 - \frac{99015}{32} e^6 - \frac{1741125}{256} e^8 - \frac{350625}{256} e^{10} - \frac{3015}{128} e^{12}\bigg) - \nu \bigg(\frac{1309361}{1680} \nonumber\\
    & + \frac{5539817}{3360} e^2 + \frac{11194711}{8960} e^4 + \frac{52850815}{10752} e^6 - \frac{34937057}{21504} e^8 - \frac{514863}{512} e^{10} - \frac{603}{64} e^{12}\bigg) + \nu^2 \bigg(\frac{1013}{4} + \frac{967}{36} e^2 \nonumber\\
    & - \frac{12503}{1152} e^4 + \frac{6274115}{2304} e^6 + \frac{4595645}{4608} e^8 - \frac{67005}{256} e^{10}\bigg)\bigg] \Lambda^{(2)}_{B,\mathrm{av}} + \bigg[-\frac{135}{4} - \frac{735}{32} e^2 - \frac{605}{32} e^4 - \frac{22295}{256} e^6 + \frac{31135}{1024} e^8 \nonumber\\
    & + \frac{805}{512} e^{10}\bigg] \Lambda^{(3)}_{A,\mathrm{av}} + \bigg[-\bigg(\frac{88273}{70} + \frac{392771}{140} e^2 - \frac{1021067}{140} e^4 + \frac{2125539}{112} e^6 + \frac{18058423}{896} e^8 - \frac{2508803}{1792} e^{10} \nonumber\\
    & - \frac{5753}{896} e^{12}\bigg) + \nu \bigg(\frac{2478}{5} - \frac{8321}{15} e^2 - \frac{125261}{30} e^4 + \frac{139121}{12} e^6 + \frac{1422131}{96} e^8 - \frac{45341}{96} e^{10} - \frac{233}{48} e^{12}\bigg)\bigg] \Sigma^{(2)}_{A,\mathrm{av}} \nonumber\\
    & - \nu \bigg(\frac{4284}{5} + \frac{25109}{30} e^2 - \frac{20111}{60} e^4 + \frac{62713}{16} e^6 + \frac{26299}{192} e^8 - \frac{19159}{96} e^{10}\bigg) \Sigma^{(2)}_{B,\mathrm{av}} \bigg\} \, , \label{eq:aTecc:14} \\
    a_{15}^\mathrm{T} =& \pi \bigg\{ \bigg[\frac{62199}{280} \theta_{y,\Lambda_A,15} - \nu \frac{12469}{20} \theta_{y,\nu\Lambda_A,15}\bigg] \Lambda^{(2)}_{A,\mathrm{av}} + \bigg[\frac{152}{35} \theta_{y,\Lambda_B,15} + \nu \frac{5139}{20} \theta_{y,\nu\Lambda_B,15}\bigg] \Lambda^{(2)}_{B,\mathrm{av}} + \frac{5528}{5} \theta_{y,\Sigma_A,15} \Sigma^{(2)}_{A,\mathrm{av}} \nonumber\\
    & - \frac{8}{15} \theta_{y,\Sigma_B,15} \Sigma^{(2)}_{B,\mathrm{av}} \bigg\} \, , \label{eq:aTecc:15}
\end{align}
\end{subequations}

\noindent The corresponding contributions to the evolution of the eccentricity, Eq.~\eqref{eq:RR_eqs_tides:e2}, are

\begin{subequations}
\label{eq:bTecc}
\begin{align}
    b_{10}^\mathrm{T} =& e^2 \bigg[\frac{2049}{10} + \frac{45143}{40} e^2 + \frac{186103}{160} e^4 + \frac{283703}{1280} e^6 + \frac{135}{32} e^8\bigg] \Lambda^{(2)}_{A,\mathrm{av}} + e^2 \bigg[\frac{306}{5} + \frac{4383}{10} e^2 + \frac{2241}{4} e^4 + \frac{4581}{32} e^6 \nonumber\\
    & + \frac{117}{32} e^8\bigg] \Lambda^{(2)}_{B,\mathrm{av}} + e^2 \sqrt{1 - e^2} \bigg[110 + \frac{7391}{40} e^2 + \frac{4057}{160} e^4 - \frac{221}{160} e^6\bigg] \Lambda^{(2)}_{A,\mathrm{av}} \, , \label{eq:bTecc:10} \\
    b_{12}^\mathrm{T} =& e^2 \bigg[-\bigg(\frac{8119}{14} - \frac{23909213}{2240} e^2 - \frac{181564903}{4480} e^4 - \frac{978752881}{35840} e^6 - \frac{34386317}{10240} e^8 - \frac{69129}{1792} e^{10}\bigg) - \nu \bigg(\frac{7517}{15} \nonumber\\
    & + \frac{965723}{80} e^2 + \frac{16085557}{480} e^4 + \frac{14078923}{640} e^6 + \frac{2249095}{768} e^8 + \frac{2565}{64} e^{10}\bigg)\bigg] \Lambda^{(2)}_{A,\mathrm{av}} + e^2 \bigg[-\bigg(\frac{6927}{70} - \frac{254567}{80} e^2 \nonumber\\
    & - \frac{4185269}{280} e^4 - \frac{11069089}{896} e^6 - \frac{18209263}{8960} e^8 - \frac{458709}{14336} e^{10}\bigg) + \nu \bigg(\frac{3837}{20} - \frac{194121}{80} e^2 - \frac{3890003}{320} e^4 - \frac{5607683}{512} e^6 \nonumber\\
    & - \frac{998717}{512} e^8 - \frac{2277}{64} e^{10}\bigg)\bigg] \Lambda^{(2)}_{B,\mathrm{av}} + e^2 \bigg[\frac{1516}{3} + \frac{202949}{30} e^2 + \frac{21875}{2} e^4 + \frac{1013453}{240} e^6 + \frac{198221}{960} e^8 \nonumber\\
    & - \frac{225}{128} e^{10}\bigg] \Sigma^{(2)}_{A,\mathrm{av}} + e^2 \bigg[-\frac{76}{3} - 369 e^2 - 1057 e^4 - \frac{6265}{8} e^6 - \frac{4347}{32} e^8 - \frac{329}{128} e^{10}\bigg] \Sigma^{(2)}_{B,\mathrm{av}} \nonumber\\
    & + e^2 \sqrt{1 - e^2} \bigg\{ \bigg[\bigg(\frac{301437}{280} + \frac{1060691}{160} e^2 + \frac{3173225}{448} e^4 + \frac{2859989}{8960} e^6 - \frac{191119}{8960} e^8\bigg) - \nu \bigg(\frac{13087}{30} + \frac{324351}{80} e^2 \nonumber\\
    & + \frac{59527}{16} e^4 + \frac{254477}{640} e^6 - \frac{33163}{1920} e^8\bigg)\bigg] \Lambda^{(2)}_{A,\mathrm{av}} + \nu \bigg(\frac{1555}{3} + \frac{15065}{16} e^2 + \frac{87575}{192} e^4 - \frac{3625}{64} e^6\bigg) \Lambda^{(2)}_{B,\mathrm{av}} + \bigg[2488 \nonumber\\
    & + \frac{11403}{2} e^2 + \frac{20749}{4} e^4 + \frac{171}{16} e^6 + \frac{1}{16} e^8\bigg] \Sigma^{(2)}_{A,\mathrm{av}} \bigg\} \, , \label{eq:bTecc:12} \\
    b_{13}^\mathrm{T} =& \pi e^2 \bigg\{ \frac{14219}{8} \theta_{e,\Lambda_A,13} \Lambda^{(2)}_{A,\mathrm{av}} + \frac{1491}{4} \theta_{e,\Lambda_B,13} \Lambda^{(2)}_{B,\mathrm{av}} \bigg\} \, , \label{eq:bTecc:13} \\
    b_{14}^\mathrm{T} =& e^2 \bigg[-\bigg(\frac{82743697}{15120} + \frac{1216333963}{17280} e^2 - \frac{2720493533}{24192} e^4 - \frac{24485741279}{46080} e^6 - \frac{344719420351}{1290240} e^8 - \frac{6451484171}{286720} e^{10} \nonumber\\
    & - \frac{9003637}{57344} e^{12}\bigg) + \nu \bigg(\frac{1827601}{210} + \frac{4490959}{140} e^2 - \frac{2726590009}{6720} e^4 - \frac{10076963933}{10752} e^6 - \frac{578882411}{1344} e^8 \nonumber\\
    & - \frac{2113698147}{57344} e^{10} - \frac{495969}{1792} e^{12}\bigg) - \nu^2 \bigg(\frac{38227}{144} - \frac{99997487}{2880} e^2 - \frac{3287046031}{11520} e^4 - \frac{43147973551}{92160} e^6 \nonumber\\
    & - \frac{9464460739}{46080} e^8 - \frac{225900263}{11520} e^{10} - \frac{6255}{32} e^{12}\bigg)\bigg] \Lambda^{(2)}_{A,\mathrm{av}} + e^2 \bigg[-\bigg(\frac{9445}{7} + \frac{1435301}{160} e^2 - \frac{130189411}{1680} e^4 \nonumber\\
    & - \frac{1044284889}{4480} e^6 - \frac{455325695}{3584} e^8 - \frac{5928020881}{430080} e^{10} - \frac{3729585}{28672} e^{12}\bigg) + \nu \bigg(\frac{4234787}{1680} + \frac{102715421}{1920} e^2 \nonumber\\
    & - \frac{1426909}{4480} e^4 - \frac{1962033783}{7168} e^6 - \frac{27436469973}{143360} e^8 - \frac{461104567}{20480} e^{10} - \frac{118927}{512} e^{12}\bigg) - \nu^2 \bigg(\frac{195761}{120} + \frac{326623}{24} e^2 \nonumber\\
    & - \frac{84553123}{1920} e^4 - \frac{2591459329}{15360} e^6 - \frac{162144067}{1536} e^8 - \frac{20565367}{1536} e^{10} - \frac{5655}{32} e^{12}\bigg)\bigg] \Lambda^{(2)}_{B,\mathrm{av}} + e^2 \bigg[\frac{2701}{12} + \frac{467425}{192} e^2 \nonumber\\
    & + \frac{137051}{32} e^4 + \frac{717151}{384} e^6 + \frac{945883}{6144} e^8 + \frac{105}{128} e^{10}\bigg] \Lambda^{(3)}_{A,\mathrm{av}} + e^2 \bigg[-\bigg(\frac{542939}{210} - \frac{12939691}{210} e^2 - \frac{585876119}{1680} e^4 \nonumber\\
    & - \frac{1005358889}{2688} e^6 - \frac{629048467}{6720} e^8 - \frac{322666189}{107520} e^{10} + \frac{143573}{7168} e^{12}\bigg) + \nu \bigg(\frac{9268}{15} - \frac{438757}{10} e^2 - \frac{4344581}{20} e^4 \nonumber\\
    & - \frac{36017293}{160} e^6 - \frac{85861273}{1440} e^8 - \frac{19321867}{11520} e^{10} + \frac{825}{32} e^{12}\bigg)\bigg] \Sigma^{(2)}_{A,\mathrm{av}} + e^2 \bigg[\bigg(\frac{1099}{15} - \frac{9062}{15} e^2 - \frac{432463}{40} e^4 \nonumber\\
    & - \frac{893949}{40} e^6 - \frac{1419607}{128} e^8 - \frac{38423}{32} e^{10} - \frac{12529}{1024} e^{12}\bigg) + \nu \bigg(\frac{86654}{45} + \frac{156453}{4} e^2 + \frac{4671832}{45} e^4 + \frac{126391667}{1440} e^6 \nonumber\\
    & + \frac{137091827}{5760} e^8 + \frac{6057949}{2880} e^{10} + \frac{2443}{96} e^{12}\bigg)\bigg] \Sigma^{(2)}_{B,\mathrm{av}} + e^2 \sqrt{1 - e^2} \bigg\{ \bigg[\bigg(\frac{652145}{63} + \frac{465567461}{3360} e^2 + \frac{28726706081}{80640} e^4 \nonumber\\
    & + \frac{265153067}{1344} e^6 + \frac{181722059}{17920} e^8 - \frac{443487}{8960} e^{10}\bigg) - \nu \bigg(\frac{402401}{112} + \frac{98244221}{1120} e^2 + \frac{3567895967}{13440} e^4 + \frac{4217364169}{26880} e^6 \nonumber\\
    & + \frac{59563387}{7168} e^8 - \frac{21617483}{107520} e^{10}\bigg) - \nu^2 \bigg(\frac{70141}{90} - \frac{49912583}{2880} e^2 - \frac{246826639}{3840} e^4 - \frac{48388043}{1280} e^6 - \frac{17478931}{5760} e^8 \nonumber\\
    & + \frac{16465}{144} e^{10}\bigg)\bigg] \Lambda^{(2)}_{A,\mathrm{av}} + \bigg[\bigg(2439 + \frac{123759}{4} e^2 + \frac{607905}{8} e^4 + \frac{3004065}{64} e^6 + \frac{207675}{32} e^8 + \frac{5805}{64} e^{10}\bigg) \nonumber\\
    & + \nu \bigg(\frac{4660647}{560} + \frac{7545683}{210} e^2 + \frac{167396401}{5376} e^4 - \frac{25931313}{3584} e^6 - \frac{14386763}{3584} e^8 - \frac{1161}{32} e^{10}\bigg) - \nu^2 \bigg(\frac{156737}{36} \nonumber\\
    & + \frac{3704137}{144} e^2 + \frac{33915575}{1152} e^4 + \frac{16762745}{2304} e^6 - \frac{621985}{768} e^8\bigg)\bigg] \Lambda^{(2)}_{B,\mathrm{av}} + \bigg[\frac{1735}{4} + \frac{60975}{64} e^2 + \frac{98345}{128} e^4 - \frac{29745}{512} e^6 \nonumber\\
    & - \frac{2515}{512} e^8\bigg] \Lambda^{(3)}_{A,\mathrm{av}} + \bigg[\bigg(\frac{3021359}{210} + \frac{22628569}{140} e^2 + \frac{47491121}{168} e^4 + \frac{88288349}{672} e^6 - \frac{217397}{128} e^8 - \frac{95765}{2688} e^{10}\bigg) \nonumber\\
    & - \nu \bigg(\frac{174674}{15} + \frac{1638601}{15} e^2 + \frac{542659}{3} e^4 + \frac{2167483}{24} e^6 + \frac{16613}{24} e^8 - \frac{1411}{48} e^{10}\bigg)\bigg] \Sigma^{(2)}_{A,\mathrm{av}} + \nu \bigg(\frac{165172}{15} \nonumber\\
    & + \frac{1894837}{60} e^2 + \frac{293335}{8} e^4 + \frac{462077}{96} e^6 - \frac{59857}{96} e^8\bigg) \Sigma^{(2)}_{B,\mathrm{av}} \bigg\} \, , \label{eq:bTecc:14} \\
    b_{15}^\mathrm{T} =& \pi e^2 \bigg\{ \bigg[\frac{728991}{280} \theta_{e,\Lambda_A,15} - \nu \frac{2826623}{240} \theta_{e,\nu\Lambda_A,15}\bigg] \Lambda^{(2)}_{A,\mathrm{av}} + \bigg[-\frac{144331}{140} \theta_{e,\Lambda_B,15} + \nu \frac{693365}{336} \theta_{e,\nu\Lambda_B,15}\bigg] \Lambda^{(2)}_{B,\mathrm{av}} \nonumber\\
    & + \frac{230774}{15} \theta_{e,\Sigma_A,15} \Sigma^{(2)}_{A,\mathrm{av}} - \frac{1528}{15} \theta_{e,\Sigma_B,15} \Sigma^{(2)}_{B,\mathrm{av}} \bigg\} \, , \label{eq:bTecc:15}
\end{align}
\end{subequations}

\noindent and those to the periastron advance, Eq.~\eqref{eq:RR_eqs_tides:k}, are

\begin{subequations}
\label{eq:kTecc}
\begin{align}
    k_{8}^\mathrm{T} =& \bigg[\frac{45}{32} + \frac{135}{64} e^2 + \frac{45}{256} e^4\bigg] \Lambda^{(2)}_{A,\mathrm{av}} \, , \label{eq:kTecc:8} \\
    k_{10}^\mathrm{T} =& \bigg[\bigg(\frac{81}{8} + \frac{2187}{32} e^2 + \frac{5373}{128} e^4 + \frac{855}{512} e^6\bigg) + \nu \bigg(\frac{75}{64} - \frac{2265}{128} e^2 - \frac{9825}{512} e^4 - \frac{615}{512} e^6\bigg)\bigg] \Lambda^{(2)}_{A,\mathrm{av}} + \nu \bigg(\frac{315}{64} + \frac{1575}{128} e^2 \nonumber\\
    & + \frac{1575}{512} e^4\bigg) \Lambda^{(2)}_{B,\mathrm{av}} + \bigg[\frac{39}{2} + \frac{135}{2} e^2 + \frac{495}{16} e^4 + \frac{75}{64} e^6\bigg] \Sigma^{(2)}_{A,\mathrm{av}} + \sqrt{1 - e^2} \bigg[\frac{9}{8} + \frac{81}{32} e^2 + \frac{9}{16} e^4\bigg] \Lambda^{(2)}_{A,\mathrm{av}} \, , \label{eq:kTecc:10} \\
    k_{12}^\mathrm{T} =& \bigg[\bigg(\frac{2061}{64} + \frac{21393}{32} e^2 + \frac{163395}{128} e^4 + \frac{371457}{1024} e^6 + \frac{28215}{4096} e^8\bigg) + \nu \bigg(\frac{7263}{256} - \frac{27015}{512} e^2 - \frac{1381617}{2048} e^4 - \frac{83097}{256} e^6 \nonumber\\
    & - \frac{18615}{2048} e^8\bigg) - \nu^2 \bigg(\frac{665}{64} + \frac{5005}{128} e^2 - \frac{56235}{512} e^4 - \frac{93035}{1024} e^6 - \frac{8815}{2048} e^8\bigg)\bigg] \Lambda^{(2)}_{A,\mathrm{av}} + \bigg[\nu \bigg(\frac{17991}{256} + \frac{256329}{512} e^2 \nonumber\\
    & + \frac{978975}{2048} e^4 + \frac{59985}{1024} e^6\bigg) - \nu^2 \bigg(\frac{945}{64} + \frac{19215}{128} e^2 + \frac{92925}{512} e^4 + \frac{29925}{1024} e^6\bigg)\bigg] \Lambda^{(2)}_{B,\mathrm{av}} + \bigg[\frac{105}{32} + \frac{1575}{128} e^2 + \frac{1575}{256} e^4 \nonumber\\
    & + \frac{525}{2048} e^6\bigg] \Lambda^{(3)}_{A,\mathrm{av}} + \bigg[\bigg(\frac{543}{4} + \frac{5763}{4} e^2 + \frac{7965}{4} e^4 + \frac{14511}{32} e^6 + \frac{4125}{512} e^8\bigg) - \nu \bigg(\frac{29}{4} + 429 e^2 + \frac{28245}{32} e^4 \nonumber\\
    & + \frac{35465}{128} e^6 + \frac{975}{128} e^8\bigg)\bigg] \Sigma^{(2)}_{A,\mathrm{av}} + \nu \bigg(\frac{289}{4} + 357 e^2 + \frac{8925}{32} e^4 + \frac{4165}{128} e^6\bigg) \Sigma^{(2)}_{B,\mathrm{av}} + \sqrt{1 - e^2} \bigg\{ \bigg[\bigg(\frac{1827}{32} + \frac{2061}{8} e^2 \nonumber\\
    & + \frac{42903}{256} e^4 + \frac{1377}{128} e^6\bigg) - \nu \bigg(\frac{309}{16} + \frac{6261}{64} e^2 + \frac{2157}{32} e^4 + 6 e^6\bigg)\bigg] \Lambda^{(2)}_{A,\mathrm{av}} + \nu \bigg(\frac{75}{16} + \frac{825}{64} e^2 + \frac{225}{32} e^4\bigg) \Lambda^{(2)}_{B,\mathrm{av}} + \bigg[\frac{45}{2} \nonumber\\
    & + \frac{315}{4} e^2 + \frac{1125}{16} e^4 + \frac{45}{8} e^6\bigg] \Sigma^{(2)}_{A,\mathrm{av}} \bigg\} \, , \label{eq:kTecc:12}
\end{align}
\end{subequations}

\noindent where, following Ref.~\cite{Morras:2025nlp}, we have introduced the enhancement functions

\begin{subequations}
\label{eq:thetaTecc}
\begin{align}
    \theta_{y,\Lambda_A,13} =& 1 + \frac{24679}{2304} e^2 + \frac{36403}{1536} e^4 + \frac{3092927}{221184} e^6 + \frac{20138917}{21233664} e^8 - \frac{242265263}{283115520} e^{10} + \ord{e^{12}} \, , \\
    \theta_{y,\Lambda_B,13} =& 1 + \frac{1689}{64} e^2 + \frac{13509}{128} e^4 + \frac{3633127}{36864} e^6 + \frac{1002787}{49152} e^8 + \frac{52964281}{117964800} e^{10} + \ord{e^{12}} \, , \\
    \theta_{y,\Lambda_A,15} =& 1 + \frac{8244257}{124398} e^2 + \frac{3122110229}{7961472} e^4 + \frac{8425780561}{15922944} e^6 + \frac{72889899301}{382150656} e^8 - \frac{321835060033}{38215065600} e^{10} + \ord{e^{12}} \, , \\
    \theta_{y,\nu\Lambda_A,15} =& 1 + \frac{14581339}{523698} e^2 + \frac{543887075}{4189584} e^4 + \frac{2873062081}{16758336} e^6 + \frac{1259854466359}{19305603072} e^8 + \frac{217583883851}{965280153600} e^{10} + \ord{e^{12}} \, , \\
    \theta_{y,\Lambda_B,15} =& 1 + \frac{519939}{1216} e^2 + \frac{22964451}{4864} e^4 + \frac{7445548135}{700416} e^6 + \frac{11520947777}{1867776} e^8 + \frac{1831512481007}{2241331200} e^{10} + \ord{e^{12}} \, , \\
    \theta_{y,\nu\Lambda_B,15} =& 1 - \frac{185047}{143892} e^2 - \frac{59491027}{767424} e^4 - \frac{5288868773}{27627264} e^6 - \frac{43436295845}{378888192} e^8 - \frac{1564634629573}{88407244800} e^{10} + \ord{e^{12}} \, , \\
    \theta_{y,\Sigma_A,15} =& 1 + \frac{27953}{2764} e^2 + \frac{1750279}{66336} e^4 + \frac{19833731}{1592064} e^6 - \frac{234543767}{50946048} e^8 - \frac{78228667513}{15283814400} e^{10} + \ord{e^{12}} \, , \\
    \theta_{y,\Sigma_B,15} =& 1 + 87 e^2 + \frac{27385}{32} e^4 + \frac{286957}{144} e^6 + \frac{10492679}{8192} e^8 + \frac{368318011}{1843200} e^{10} + \ord{e^{12}} \, , \\
    \theta_{e,\Lambda_A,13} =& 1 + \frac{549103}{71095} e^2 + \frac{174556741}{13650240} e^4 + \frac{527873381}{81901440} e^6 + \frac{15318454303}{26208460800} e^8 - \frac{10560207337}{78625382400} e^{10} + \ord{e^{12}} \, , \\
    \theta_{e,\Lambda_B,13} =& 1 + \frac{30789}{2485} e^2 + \frac{5007661}{159040} e^4 + \frac{121605541}{5725440} e^6 + \frac{446499763}{130867200} e^8 + \frac{27852151}{458035200} e^{10} + \ord{e^{12}} \, , \\
    \theta_{e,\Lambda_A,15} =& 1 + \frac{797698129}{15551808} e^2 + \frac{1907561477}{7775904} e^4 + \frac{7471204533529}{26873524224} e^6 + \frac{8150141836301}{89578414080} e^8 \nonumber\\
    & + \frac{23051561741023}{20638866604032} e^{10} + \ord{e^{12}} \, , \\
    \theta_{e,\nu\Lambda_A,15} =& 1 + \frac{385730049}{22612984} e^2 + \frac{219572776067}{3798981312} e^4 + \frac{5522175308437}{91175551488} e^6 + \frac{16212773766297}{810449346560} e^8 \nonumber\\
    & + \frac{190734337323941}{250081512652800} e^{10} + \ord{e^{12}} \, , \\
    \theta_{e,\Lambda_B,15} =& 1 - \frac{20724175}{1154648} e^2 - \frac{1301470649}{6927888} e^4 - \frac{449078422319}{1330154496} e^6 - \frac{2809221442669}{17735393280} e^8 \nonumber\\
    & - \frac{22149413239687}{1276948316160} e^{10} + \ord{e^{12}} \, , \\
    \theta_{e,\nu\Lambda_B,15} =& 1 - \frac{214893997}{13867300} e^2 - \frac{81817497077}{665630400} e^4 - \frac{1552171760033}{7987564800} e^6 - \frac{22417919049971}{255602073600} e^8 \nonumber\\
    & - \frac{1599730171037041}{153361244160000} e^{10} + \ord{e^{12}} \, , \\
    \theta_{e,\Sigma_A,15} =& 1 + \frac{2938101}{461548} e^2 + \frac{223866919}{22154304} e^4 + \frac{816033053}{531703296} e^6 - \frac{105984782861}{42536263680} e^8 - \frac{3074077668623}{2552175820800} e^{10} + \ord{e^{12}} \, , \\
    \theta_{e,\Sigma_B,15} =& 1 + \frac{39095}{1528} e^2 + \frac{2434063}{18336} e^4 + \frac{354168539}{1760256} e^6 + \frac{6621836617}{70410240} e^8 + \frac{32059415041}{2816409600} e^{10} + \ord{e^{12}} \, .
\end{align}
\end{subequations}
\end{widetext}

The coefficients of even PN order, which originate from the instantaneous part of the fluxes and from the conservative dynamics, are exact in the eccentricity (no expansion in $e$ is involved). In contrast, the half-integer PN coefficients $a^\mathrm{T}_{13}$, $a^\mathrm{T}_{15}$, $b^\mathrm{T}_{13}$ and $b^\mathrm{T}_{15}$, which are proportional to $\pi$ and arise from the tail contributions, are only available as an expansion in eccentricity. Following Ref.~\cite{Morras:2025nlp}, we factor each of them into a rational prefactor and an enhancement function $\theta$, normalized to unity in the quasi-circular limit, whose subscripts identify the series ($y$ or $e^2$), the monomial it multiplies, and the PN index. Of each expansion, we retain six terms, i.e., all contributions through relative order $e^{10}$ with respect to the leading term of each coefficient. Further orders, up to relative $e^{14}$, are available in the ancillary files of Refs.~\cite{Henry:2025uta,Henry:2026bqh}.

Note that the tidal corrections from Refs.~\cite{Henry:2025uta,Henry:2026bqh} implemented here treat the tides as adiabatic. For eccentric orbits, the repeated pericentre passages are expected to excite the stellar oscillation modes, so dynamical tides should play a qualitative role that the present treatment does not capture~\cite{Henry:2025uta}. We also include only the secular evolution of the orbital elements. The post-adiabatic (oscillatory) corrections provided in Ref.~\cite{Henry:2025uta} are not used. 

Since Refs.~\cite{Henry:2025uta,Henry:2026bqh} consider non-spinning binaries, the 6.5PN spin-tidal coefficient $a^\mathrm{ST,QC}_{13}$ of Refs.~\cite{Abdelsalhin:2018reg,Castro:2022mpw} is retained in its quasi-circular form, no spin-tidal contributions to $\D e^2$ or $k$ are included, and the rotational tidal Love numbers are set to zero (see Appendix~\ref{sec:appendix:QUR}). None of this affects the present study, whose configurations are all non-spinning, so that these terms vanish identically. Finally, we include eccentric tidal corrections in the phasing only.

\section{Neutron-star quasi-universal relations}
\label{sec:appendix:QUR}

Besides the quadrupolar tidal deformability $\Lambda^{(2)}_i$, the model needs the octupolar and current-quadrupolar deformabilities $\Lambda^{(3)}_i$ and $\Sigma^{(2)}_i$. It also needs the spin-induced quadrupole and octupole parameters $q_i$ and $o_i$, which enter the spin-spin and cubic-in-spin sectors and are equal to unity for black holes, and not to be confused with the mass ratio $q$. Rather than computing each of these from the EOS, we obtain them from $\Lambda^{(2)}_i$ using the EOS-insensitive quasi-universal relations. Each takes the form of a polynomial fit of a quantity $Y_i$ in an argument $x_i$,
\begin{equation}
    Y_i = \sum_{n=0}^{N} c_n \, x_i^{\,n} \, ,   \qquad N = 4 \text{ or } 6 \, ,
    \label{eq:QUR_form}
\end{equation}

\noindent with $Y_i$, $x_i$ and the coefficients $c_n$ listed in Table~\ref{tab:QUR}. We use the $I$-Love-$Q$ relations of Ref.~\cite{Yagi:2016bkt} for $q_i$, and their multipole no-hair relation for $o_i$. For $\Lambda^{(3)}_i$, $\Sigma^{(2)}_i$ and the compactness $\mathcal{C}_i = m_i/R_i$ we use the updated multipole-Love relations of Ref.~\cite{Pradhan:2022rxs}, which were calibrated on a large set of EOSs. Following the convention of Refs.~\cite{Henry:2020ski,Abdelsalhin:2018reg,JimenezForteza:2018rwr}, $\Sigma^{(2)}_i$ is negative for an irrotational fluid. No fit is currently available for the rotational tidal Love numbers $\Lambda^{(23)}_i$, $\Lambda^{(32)}_i$, $\Sigma^{(23)}_i$, and $\Sigma^{(32)}_i$ in the convention of Ref.~\cite{Abdelsalhin:2018reg}, so these are set to zero.

\begin{table*}[!htbp]
    \centering
    \begin{tabular}{c|c|ccccccc}
    $Y$ & $x$ & $c_0$ & $c_1$ & $c_2$ & $c_3$ & $c_4$ & $c_5$ & $c_6$ \\
    \hline
    $\ln q_i$                   & $\ln \Lambda^{(2)}$ & $0.1940$  & $0.09163$  & $0.04812$   & $-4.283\times 10^{-3}$ & $1.245\times 10^{-4}$  & & \\
    $\ln o_i$                   & $\ln q_i$           & $3.131\times 10^{-3}$ & $2.071$ & $-0.7152$ & $0.2458$ & $-0.03309$ & & \\
    $\ln \Lambda^{(3)}$         & $\ln \Lambda^{(2)}$ & $-1.163$  & $0.9442$   & $0.2492$    & $-0.08170$ & $0.01374$ & $-1.117\times 10^{-3}$ & $3.494\times 10^{-5}$ \\
    $\ln \lvert\Sigma^{(2)}\rvert$ & $\ln \Lambda^{(2)}$ & $-2.019$  & $0.4821$   & $7.609\times 10^{-4}$ & $4.096\times 10^{-3}$ & $-5.030\times 10^{-4}$ & $2.643\times 10^{-5}$ & $-5.192\times 10^{-7}$ \\
    $\mathcal{C}$               & $\ln \Lambda^{(2)}$ & $0.3663$  & $-0.03727$ & $-2.243\times 10^{-3}$ & $1.941\times 10^{-3}$ & $-4.491\times 10^{-4}$ & $4.463\times 10^{-5}$ & $-1.606\times 10^{-6}$ \\
    \end{tabular}
    \caption{Fitted quantity $Y$, argument $x$ and coefficients $c_n$ of the quasi-universal relations of Eq.~\eqref{eq:QUR_form}, from Ref.~\cite{Yagi:2016bkt} ($q$, $o$) and Ref.~\cite{Pradhan:2022rxs} ($\Lambda^{(3)}$, $\Sigma^{(2)}$, $\mathcal{C}$). Note that the spin-induced octupole is fitted in $\ln q$ rather than in $\ln \Lambda^{(2)}$, and that the compactness is the one quantity fitted directly rather than through its logarithm.}
    \label{tab:QUR}
\end{table*}

\section{Carbon footprint}
\label{section:carbon_footprint}
As in our previous works~\cite{Markin:2023fxx,Markin:2025oeo,Neuweiler:2025klw,Markin:2026eyc}, we estimate the energy required to perform the simulations that support this paper, and the corresponding greenhouse gas (GHG) emissions induced by consumption of electricity from the national grid, and list them in Table~\ref{table:carbon_footprint}. The radiative forcing potential of these emissions is known to be responsible for anthropogenically induced climate change~\cite{Oreskes_2004,Doran_2009,Archer:2009,Unger:2010,Cook_2013,Cook_2016,Fahey:2017,Lynas_2021,Myers_2021}.
We stress that these emissions can be avoided by switching to renewable energy sources, which is done for the Emmy supercomputer NHR-Nord@Göttingen~\cite{NHRGoettingenPUE}, on which we ran our simulations.

To estimate the energy consumption and the corresponding induced GHG emissions, we employ the \textsc{calcium}~\cite{calcium} software tool. \textsc{calcium} automatically collects and calculates the energy from CPU utilization via thermal design power of the CPU model. To convert the energy to the induced GHG emissions, \textsc{calcium} uses the national electricity grid carbon intensity values~\cite{owid_electricity_mix_2023}.
To account for the energy consumption of non-CPU components of the compute nodes, we adopt a multiplicative factor of $1.3$, calibrated on the power metrics provided by NHR@Göttingen. For the power usage effectiveness, we adopt a value of $1.03$ for Emmy~\cite{NHRGoettingenPUE}.

\begin{table}[!htbp]
    \centering
    \begin{tabular}{l|c|c}
        & Energy & Induced grid emissions,\\
        & MWh & t\ch{CO2}e \\
       \hline
Evolution R192 & 13.8 & 4.5 \\
Evolution R144 & 5.4 & 1.8 \\
Evolution R96 & 1.7 & 0.6 \\
Evolution GRMHD R192 & 3.4 & 1.1 \\
Test runs & 8.9 & 2.9 \\
\hline
Total & 33.2 & 11.0 \\
    \end{tabular}
    \caption{Energy consumption for each step of this study and the corresponding estimated induced \ch{CO2}e emissions.}
    \label{table:carbon_footprint}
\end{table}

\bibliography{main}

\end{document}